**Thinker is the Clear Winner**
**Uyghur Proverb**

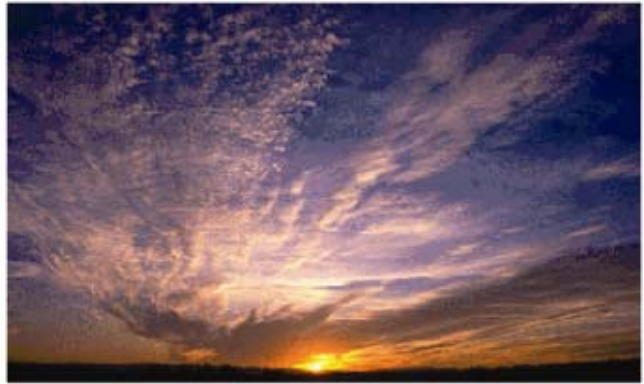

# MICROWAVE TECHNOLOGIES

## Determination of Magnetic and Dielectric Materials Microwave Properties

Mahmut Obol, PhD

*Disclaimer: The techniques in this summary; each of them works very well for the suitable subjects and they are well practiced by real time measurements of microwave materials such as magnetic and dielectrics. However, there will be no responsibility applicable for inappropriate applications and relevant losses. June 16, 2009.*

**Abstract**

In this research summary, four different microwave measurement techniques are presented for materials characterization. They are the following. 1) Rectangular waveguide measurement technique for normal microwave materials microwave properties such as permeability and permittivity. This technique removed guess parameter and dispersive effect issues of the old waveguide measurement techniques forty years history. As such it projects a new route for determination of any microwave materials magnetic and dielectric properties without using any guesses. 2) Coaxial probe measurement technique for the liquid and biological tissues dielectric permittivity. This coaxial probe technique has an advantage which is to attain the highest reflected signal from the coaxial probe tip, so that it is a fast and very sensitive technique to differentiate lossy materials dielectric permittivity. This technique could be useful non destructive detections for tumors in hospital and non destructive detections for chemical liquids as well. 3) A microstripline measurement technique for oxides microwave measurement at low frequency spectra where the waveguide technique becomes robot and cumbersome. 4) A new methodology is presented for the rectangular waveguide technique to determine microwave metamaterials refractive index, permeability and permittivity using the rectangular waveguide. In summary, the presented techniques are capable enough to determine magnetic and non magnetic solid state materials, liquids, powders and biological tissues microwave properties from the broad microwave frequencies spectra between 1 GHz to 50 GHz. Appropriate advices and useful micro codes of measurement processes will be provided to interested agent, based on their appropriate requests.

Lastly, a special ferrite's microwave properties are also presented in this research summary since the negative refractive index from the insulator materials such as ferrite is an interesting subject.

# CONTENTS

# Microwave Materials Permeability and Permittivity Measurements by T/R Technique in Rectangular Waveguide

*Abstract* — There is a huge demand to accurately determine the magneto-electrical properties of solids, and particles in the nano sized regime due to the modern IC technology revolution and biomedical application science. In this paper, one presents a microwave waveguide measurement technique for complex permeability and permittivity of expensive nano sized magnetic powder materials. In the measurement process, Agilent's 8510C vector network analyzer was used to have a standard TRL calibration for empty space inside the waveguides. In order to maintain the recommended insertion phase range, a very thin prepared sample was loaded inside the calibrated waveguide. The loaded material's magnetic and dielectric effects were also considered into the cutoff wavelength calculation of the propagation constant of the TE$_{10}$ wave from the geometrical dimensions of the waveguides. These considerations make the measured permeability and permittivity more reliable than commonly used techniques. However, at this time this technique is capable enough to determine microwave properties for the thick samples up to 0.5cm.

## I. INTRODUCTION

Nicolson and Ross [1] developed a broadband simultaneous measurement technique of transmission and reflection (T/R) by using forward and backward energy scattering in coaxial transmission line. Weir [2] extended it into waveguide transmission line by using a Hewlett-Packard vector network analyzer. J. Baker-Jarvis [3] has solved the phase ambiguity problem in [2] by using a reasonable guessing parameter to detect reasonable permittivity in relatively thick samples, in which the samples can be both lossless and lossy materials. The other attempts were also studied in [4] to remove inaccurate reflection peaks in complex permittivity measurements. Since then the transmission and reflection technique was also used in the free space measurement technique [5, 6]. The S-matrix analysis was also deployed for the coaxial [7] and rectangular [8, 9] waveguide techniques to determine the complex permeability and complex permittivity of specific materials. These measurements of the complex permeability and permittivity are very reliable. However, the minimum diameter requirement of Gaussian beams in free space measurement always requires a wavelength for operating at a central frequency. In order to avoid diffraction errors from target materials, the surface diameter of the target material was at least three times larger than the diameter of the Gaussian beam. This implies that in order to determine a material's electro-magnetic properties at lower frequencies, a large amount of material is needed in the measurement process. Due to the relatively expensive and nominal nature of nano materials, a cost effective measurement technique requires a minimal sample size. For example, coaxial transmission line technique [10] was deployed to measure the complex permeability and permittivity of nano materials. The waveguide technique also fits the requirements for this and delivers very accurate results. It is obvious that each waveguide has a limited frequency band; however, they do not suffer radiation losses like free space measurements except for the attenuation losses of specific modes in the waveguide. In order to remove any unwanted losses from the waveguide, we applied a standard TRL calibration [8] technique to achieve a zero reference plane for the measurements inside the waveguide. After a great number of trials using TRL calibration, we noticed that loaded material's permittivity and permeability effects into the cutoff wavelengths from the free space waveguide need to be accounted for. Nevertheless, we have not seen this phenomenon explicitly expressed in any relevant works. In addition, the loaded lossy material's thickness should be as thinner as possible for lower frequency measurements in order to maintain the recommended insertion phase regime by Agilent. The transparent ticker sample is hold to transparent signal, providing very accurate S-parameters and no phase ambiguity. It was also reported [3] that thinner sample loading technique was a source of errors due to uncertainty in reference plane positions. However, we noted that using electrical delay function of VNA, one could eliminate those errors. To account for the thin and thick loaded sample, one presents a modified reflection and transmission formulation for in-waveguide measurements. Also presented is a modified permeability and permittivity formulation from the modified propagation constant of the loaded waveguide. The studies show that these modifications are necessary and known data is presented to confirm the accuracy of the measurement technique. The derived permeability and permittivity data is very reliable and not effected by the scattering voltage ratios of the vector network analyzer.

## II. THEORY AND MEASUREMENT

A propagating electromagnetic wave inside the waveguide is being reflected, $S_{11}$, and transmitted, $S_{21}$, by the loaded material. A diagram of this setup can be seen in Fig.1. The known electric and magnetic polarization of propagating waves in waveguides is very useful in analyzing the physical properties of certain materials. It is also very useful in determining the dielectric and magnetic complex permeability and permittivity of these materials.

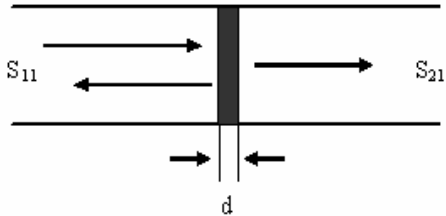

Fig.1 Schematic diagram of powder sample in waveguide

In this waveguide measurement technique, the standard TRL calibration applies to the zero reference plane findings. The zero reference planes were realized by the typical quarter wavelength difference ($l$) between *thru* and *line* in air. The recommended insertion phase ranges from $20^0$ to $160^0$ are retained by proper TRL calibration. In order for the insertion phase contributions from air to be removed from the actual transmission line for the loaded material measurements, the reasonable thin and thick prepared target materials were loaded inside the waveguide and mounted onto the zero reference planeside. The modified S parameters are as follows:

$$\widetilde{S}_{11} = S_{11}e^{j(0\times\sqrt{k_0^2-k_c^2})}$$
$$\widetilde{S}_{21} = S_{21}e^{j(l-d)\times\sqrt{k_0^2-k_c^2}} \qquad (1)$$

Return losses of less than -50 dB from the air inside the waveguide are easily achieved using the calibration techniques described. This enables us to neglect any unwanted reflections from the inner walls of the waveguide when analyzing the S parameters. The reflection and transmission by the scattering parameters inside the waveguide, in which the transmission and reflection may be resembled by free space formulations, can now be presented as follows:

$$\Gamma = K \pm \sqrt{K^2 - 1}$$
$$K = \frac{\widetilde{S}_{11}^2 - \widetilde{S}_{21}^2 + 1}{2\widetilde{S}_{11}} \qquad (2)$$
$$T = \frac{\widetilde{S}_{11} + \widetilde{S}_{21} - \Gamma}{1 - (\widetilde{S}_{11} + \widetilde{S}_{21})\Gamma}$$

The transmission coefficient through the material may also be written as follows: $T = e^{-\gamma d} = e^{-(\alpha + j\beta)d}$. The propagation constant through the material inside the waveguides can be derived to be:

$$\gamma_{TE_{10}} = \frac{\ln(\frac{1}{|T|})}{d} + j\left(\frac{2\pi n - \varphi_T}{d}\right) \quad (3)$$

Normally, a sample thickness of less than one quarter wavelength is desirable in this calibration, because it will make n = 0. In order to achieve our goal and derive the complex permeability and permittivity for the loaded material inside the waveguide, we must determine the propagation constant through the materials in the waveguide. To achieve this one must solve Maxwell's equations with respect to $E_y$ for the $TE_{10}$ mode as seen in Fig.2.

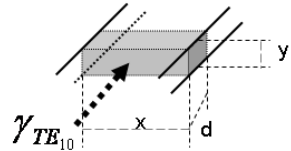

Fig.2 Propagating $TE_{10}$ wave inside waveguide and the loaded material

$$(\frac{\partial}{\partial x'^2} + \frac{\partial}{\partial y'^2} + \beta^2)E_y = 0 \qquad (4)$$

Where $x' = x\frac{1}{\sqrt{\mu\varepsilon}}$ and $y' = y\frac{1}{\sqrt{\mu\varepsilon}}$.

Solving Maxwell's equation leads to the following:

$$E_y = C\sin(\beta_x x')\cos(\beta_y y') \qquad (5)$$

C is a constant to be determined from boundary conditions. The boundary condition tells us that the propagation constant components may be presented as follows: $\gamma_0 = \frac{2\pi}{\lambda_0}$, $\beta_x = \frac{n\pi}{a}\sqrt{\mu\varepsilon}$ and $\beta_y = \frac{m\pi}{b}\sqrt{\mu\varepsilon}$.

This results in the following relationship for the total propagation constant through the material inside the waveguide: $\gamma^2 = \gamma_0^2 - \beta_x^2 - \beta_y^2$. The propagation constant of the $TE_{10}$ can subsequently be written as follows:

$$\gamma_{TE_{10}} = j2\pi\sqrt{\left(\frac{1}{\lambda_0}\right)^2 - \left(\frac{1}{2a}\right)^2} \cdot \sqrt{\mu\varepsilon} = j\gamma_{TE_{10}}^0\frac{\mu}{\eta}$$

$$\gamma_{TE_{10}}^0 = 2\pi\sqrt{\left(\frac{1}{\lambda_0}\right)^2 - \left(\frac{1}{2a}\right)^2} \qquad (6)$$

The complex permeability and permittivity associated with the propagation constant are:

$$\mu = \frac{\eta\gamma_{TE_{10}}}{j\gamma_{TE_{10}}^0} = -j\left(\frac{1+\Gamma}{1-\Gamma}\right)\left(\frac{1}{2\pi d}\right)\left(\frac{\ln(\frac{1}{|T|}) + j(2\pi n - \varphi_T)}{\sqrt{\left(\frac{1}{\lambda_0}\right)^2 - \left(\frac{1}{2a}\right)^2}}\right) \quad (7)$$

In our waveguide measurement technique the propagating wave inside the waveguide was assumed to be the $TE_{10}$ mode. This implies that the propagating wave detects the permeability directly. However, the permittivity is detected indirectly and can be derived as follows:

$$\varepsilon = \frac{\mu}{Z_{TE}^n} = \frac{\mu}{\left(Z_{TE}^n\right)^2\left(1 - \frac{\lambda_0^2}{4a^2}\right)^{-1}} = \frac{\mu}{\eta^2}\left(\lambda_0^2\left(\frac{1}{\lambda_0^2} - \frac{1}{4a^2}\right)\right)$$

$$(8)$$

Where $Z_{TE}^n = \frac{Z_{TE}^{load}}{Z_{TE}^{air}} = \frac{1+\Gamma}{1-\Gamma} = \eta$

$$\varepsilon = -j\left(\frac{c}{f}\right)^2\left(\frac{1-\Gamma}{1+\Gamma}\right)\left(\frac{1}{2\pi d}\right)\left(\ln\frac{1}{|T|} + j(2\pi n - \varphi_T)\right)\left(\sqrt{\left(\frac{1}{\lambda_0}\right)^2 - \left(\frac{1}{2a}\right)^2}\right)$$

$$(9)$$

The above permeability and permittivity equations can now be re-written by following configurations too.

$$\mu = -jZ_{eq}n_{eq} \qquad (10)$$

$$\varepsilon = -j\frac{n_{eq}}{Z_{eq}} \qquad (11)$$

Where equivalent impedance inside rectangular waveguide, and equivalent refractive index inside rectangular waveguide are the follows.

$$Z_{eq} = \eta\left(1 - \left(\frac{\lambda_0}{2a}\right)^2\right)^{-\frac{1}{2}} \qquad (12)$$

$$n_{eq} = \frac{\gamma_{TE_{10}}}{\gamma_0} = \left(\frac{2\pi}{\lambda_0}\right)\gamma_{TE_{10}} = \tilde{n}\left(1 - \left(\frac{\lambda_0}{2a}\right)^2\right)^{\frac{1}{2}} \quad (13)$$

$$\eta = \frac{1+\Gamma}{1-\Gamma} \text{ and } \tilde{n} = \kappa + jn$$

It is also possible to define the medium impedance inside waveguide by follows where one deployed simple algebra between the transmission line matrixes A, Z, and S.

$$\eta = \frac{2\sinh(\gamma_{TE_{10}}d)}{\left(\frac{(1-S_{11})^2}{S_{21}} - S_{21}\right)} \qquad (14)$$

The permeability and permittivity inside waveguide can also be figure out by following formats too.

$$\mu = -jZ_{eq}n_{eq} = -j\eta\tilde{m} = \mu_r \qquad (15)$$

$$\varepsilon = -j\frac{n_{eq}}{Z_{eq}} = -j\frac{\tilde{n}}{\eta}\left(1 - \left(\frac{\lambda_0}{2a}\right)^2\right) \qquad (16)$$

The equation (16) firmly shows that the effective permittivity inside waveguide is $\varepsilon = \varepsilon_r(1 - \left(\frac{f_c}{f}\right)^2)$.

The $f_c = \frac{c}{2a\sqrt{\varepsilon_r}}$ is the cutoff frequency of medium loaded waveguide. So that relative permittivity of the medium inside rectangular waveguide is to the follows:

$$\varepsilon_r = \varepsilon + \left(\frac{f_c^{air}}{f}\right)^2 \qquad (17)$$

However, loaded medium's relative and effective permeability inside waveguide are equal and same, and they don't have dispersive effect issues from the rectangular waveguide (see equation (15). The equations (15) and (17) are used to calculate the complex relative permeability and relative permittivity of samples inside the waveguide. Also, the $f_c^{air} = \frac{c}{2a}$ is the cutoff frequency of empty waveguide. In order to validate this technique, the simultaneous complex permeability and permittivity measurement was presented by using this technique for the known material YIG and Phenyloxide.

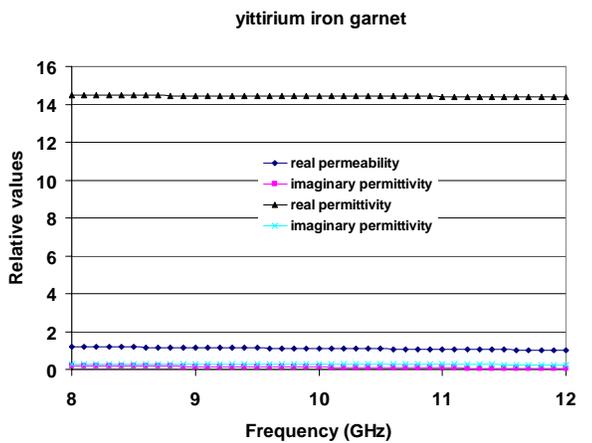

Fig.3 Complex permeability and permittivity of YIG, sample thickness 0.5 mm

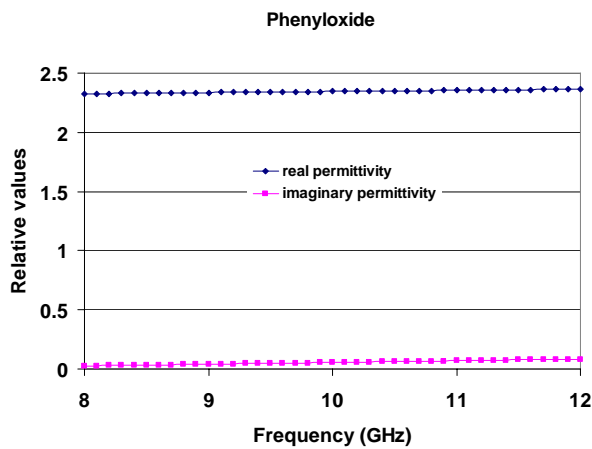

Fig.4 Complex permittivity of Phenyloxide, sample thickness 0.5 cm

The available samples of YIG and Phenyloxide were tested at sample X-band, and the sample thickness is 0.5 mm of YIG and 0.5 cm of phenyloxide and the shim thickness of TRL is 9.54mm. The YIG samples permittivity was recorded as $\varepsilon_r \cong 14.4 - j10^{-4}$ by pacific microwave ceramics brochure.

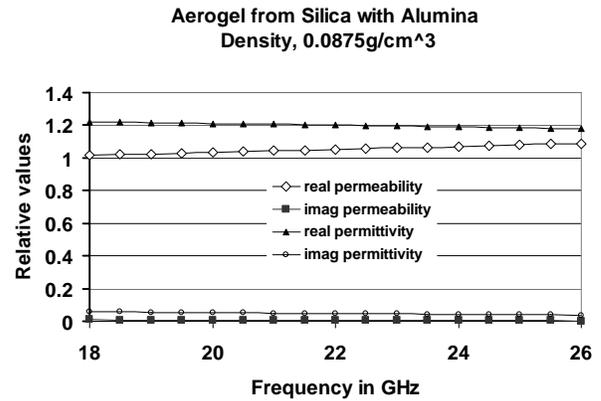

Fig.5 nano porous aerogel materials permittivity measurements, sample thickness 0.452 cm

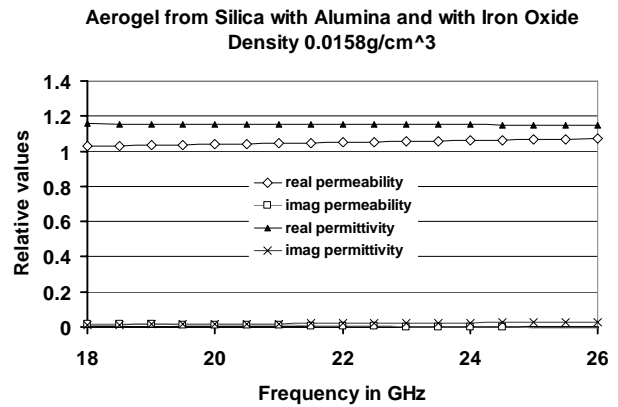

Fig. 6 iron substituted nano porous aerogel materials permittivity measurements, 0.452 cm

Looking at Figs.3, 4, 5, and 6 one is able to see that this method will not have the divergence problem like Weir; also one does not need initial guess for the permittivity determination like Baker-Jarvis. The need to obtain an appropriate reflection coefficient is crucial in determining accurate permittivity measurements. The excited ferrites permittivity and permeability determination, this method could be useful, but it is needed to be clarified some classical concepts. This part of determination will be presented in resonate mediums permittivity and permeability determination section. Now one also presents the measured complex permeability and permittivity of nano ferrite materials of magnetite using this waveguide technique.

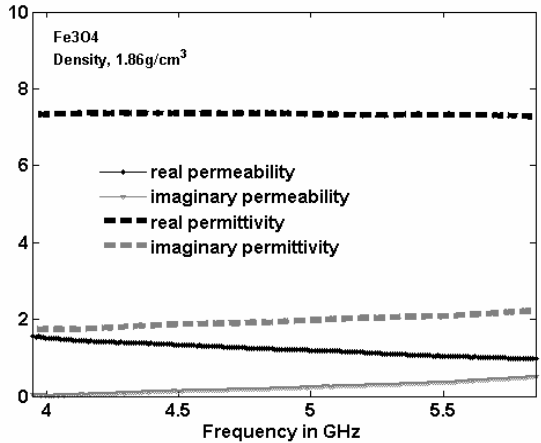

Fig.7 Complex permeability and permittivity measurements of nano spinel ferrite of magnetite

One can recall previous studies using waveguide measurements where cutoff wavelengths were not accounted for in the permeability and permittivity measurements. The technique proposed in this paper demonstrates this techniques ability to successfully account for the effect of the cutoff frequencies and obtain the potential permeability and permittivity as shown above.

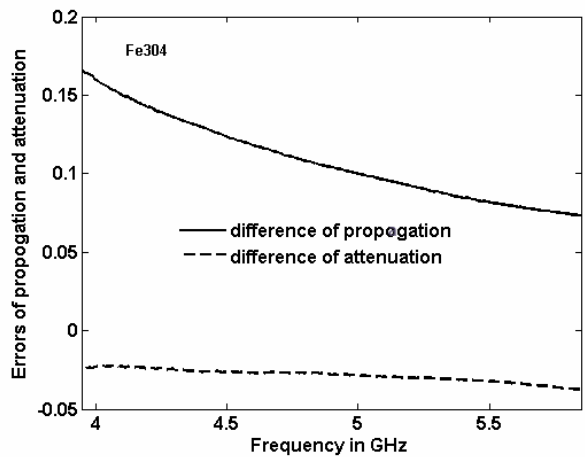

Fig.8 the cutoff wavelengths over contributions to the propagation and attenuation

## III. Conclusion

Compared to previously published data, the data obtained using the new proposed technique shall be superior by any means. This is because this technique successfully avoided the phase ambiguity and reference plane uncertainties in the measurement process. Through the analysis one can have noticed that the recommended phase can be maintained for extreme high permeability and permittivity materials by loading thinner samples. Other than that this technique even cable to measure sample thicknesses up to half centimeter. It implies that this technique is now good for thin and thick samples. It also implies that the measurement could be very successful for films such as carbon nanotubes, due to the fact that the partial insertion phase region is able to prove the reflection and transmission theories. Lastly, this technique claims that it has no guess and dispersive issues for any microwave measurements by waveguide.

### Acknowledgement

The authors wish to acknowledge the support of US Army contract, National Ground Intelligence Center.

# Coaxial Probe Technique for Microwave Characterizations of Liquid and Biological Tissues

*Abstract*— A key advantage of using a coaxial probe for microwave characterization of biological media is the non-invasive nature of the technique. Coaxial probes are being used extensively for the complex permittivity measurements of materials in the microwave region. Usually, these types of measurements require electromagnetic full-wave analysis or a calibrated reflection coefficient, $S_{11}$, of the material being tested. In this paper, a new coaxial probe technique is presented, which features the microwave characterization of biological tissues based on the calibrated reflection coefficient $S_{11}$ of a known dielectric material. Using this technique, which simply requires distilled water as a reference material, the complex permittivity of normal tissues from animals and both normal and cancerous tissues from human bodies were measured over a broadband microwave region. The biological tissue measurements show that the complex permittivity measured by this method is in concurrence with other investigations. It is a cost effective, non-invasive and practical technique, which may make it useful for diagnostic and therapeutic biomedical applications in which microwave permittivity is considered.

*Index Terms*—complex permittivity, water, biological tissue, and coaxial probe

## I. INTRODUCTION

Biological substances have been studied extensively using the coaxial probe technique, and the use of known dielectric materials as a reference in the calibration of coaxial probe measurements is common. Various problems related to coaxial probe measurements have been studied and related improvements were proposed specific to each case [1, 2, 3, 4, 5, 6, and 7]. The specific contributing papers are as follows. First, an error correction method was proposed for correcting errors from liquid reference cases [1]. Second, the temperature sensitivity of coaxial probes was studied for *in-vivo* applications [2]. Third, a through reflection and evanescent mode analysis were made for cases with air gaps between coaxial probe and test materials [3]. Fourth, an admittance model for the coaxial probe was carried out by numerical calculations to overcome the unknown coaxial probe radiations [4]. Fifth, an improved calibration technique for coaxial probes, as well as its applications for finite and infinite half space slabs, was studied [5]. Sixth, ever since this technique has been applied, the methods that involve the TEM mode propagating through a coaxial probe, as well as electromagnetic full-wave analysis and numerous rigorous mathematical modeling elsewhere, have been used to detect the permittivity of test materials by coaxial probe methods [6]. Lastly, besides the previously mentioned procedures, some other calibration techniques for coaxial probes, such as bilinear calibration, were also reported elsewhere [7]. As such, the VNA based, non-destructive coaxial probe applications were extensively employed for detecting permittivity worldwide. These methods consist of measured calibrated reflection coefficients and specific software for numerical calculations in order to detect the dielectric properties of specific materials. It is obvious that some of the above techniques have application ranges that are narrow and time consuming due to large scale mathematical modeling. In most cases, the coaxial probe technique would be significantly more reliable if a higher level of reflection was retained from the calibrated plane of the probe tip where the tested material is placed. This is because the lesser the reflection from the probe tip, the greater the error source of phase from the less reflected signal of the probe tip. Usually, the coaxial probe measurement technique is considered to be less accurate and less repeatable in real measurements that involve radiation from the coaxial probe tip to air. Bearing these factors in mind, a simple circuit model for the coaxial probe is presented, in which the largest signal content shall be reflected back from the load to the coaxial probe tip. Based on this simple circuit model, formulas were developed that are capable of producing repeatable complex microwave permittivity data for biological tissue samples. This is a fast detecting technique for real time measurements and it differentiates between the dielectric properties of various biological and liquid materials while showing reasonable data. Previously, the higher order mode excitations from the coaxial probe geometry such as $TM_{01}$, $TM_{02}$, and $TE_{01}$ were considered in coaxial probe measurements. For example, Baker-Jarvis measurements were made considering the highest order modes possible [3].

However, in this study, a coaxial probe with an outer diameter as small as 1.45 mm was implemented. The deployed operating frequency spectrum, from 2 to 18 GHz, implies that the potential higher order modes in this coaxial probe may be less critical in the computation of the dielectrics using such a smaller diameter probe. This factor greatly simplifies the approach presented here. The attempt to use the larger diameter coaxial probes for detecting the permittivity of test materials by using this technique may reduce the reliability of this technique, because it may demand one to consider the potential evanescent modes excitements in the coaxial probe measurement process. For some other cases, such as if one is interested in measuring small samples by using a coaxial probe, it will require the consideration of wave distribution on the surface of the sample, as well as penetration distances through the sample [8]. In this paper, a technique is presented for measuring complex permittivity of the materials such as biological tissues. It is not necessary to know the wave's actual penetration distances for the material that is examined.

## II. CIRCUIT AND THEORY

In using a coaxial probe technique, the examined material needs to be placed on the probe tip during the measurement process. The greatest challenge of the coaxial probe measurement may be retaining a sufficient reflected signal from the coaxial probe tip. Since the measured dielectric properties of the examined materials are based on their interaction with electromagnetic waves, an increase in electromagnetic wave radiating from the coaxial probe makes it more difficult to obtain a reliable phase from the reflected signal. This will affect the process of obtaining accurate dielectric properties of the target material. The VNA (see Fig.1) used in this study is incorporated with the usual coaxial probe for the dielectric measurement of biological tissues. A 50 ohm match load for the conventional SOL (short, open, load) was arranged for the calibration procedure, in which the reflections from the coaxial cable onto the reference plane of the probe tip could be fully removed.

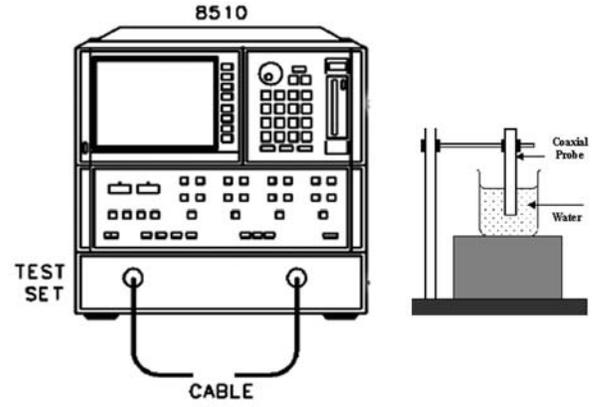

Fig.1 A sketch of the Agilent's 8510C VNA and a 2.4mm coaxial probe can be seen on the right.

Now suppose an ideal load $\varepsilon_m$ is placed between the 50 ohm transmission lines and air (see Fig.2). The circuit model below was implemented to measure the dielectric constants of distilled water and biological substances in the real experiment and measurements.

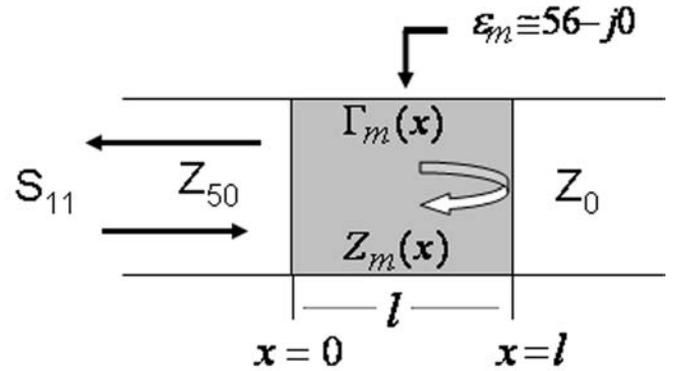

Fig.2 Circuit of an open-ended coaxial probe with an ideal load material, $\varepsilon_m \cong 56 - j0$.

In the real measurement portion of this paper, it should be understood that the materials to be measured were placed on the ideal load (see Fig.2). The formulations for the complex permittivity derivations based on the circuit model as seen in Fig. 2 will now be presented. The voltage and current should be continuous at $x = 0$ and $x = l$ according to the matched boundary conditions of the transmission lines. Also, the wave propagation constant through the load may be defined as $\gamma = \alpha + j\beta$. Normally, it is relatively easy to record the $S_{11}$ parameter at the terminal, $x = 0$, but it is not as simple to determine the reflection coefficient $\Gamma(x)$ through the actual load between $x = 0$ and $x = l$ ranges. Thus, various electromagnetic modeling and calibration methods are required to determine the realistic microwave complex permittivity of the testing materials. Let the definition for reflection coefficients be expressed as $\Gamma_m = \Gamma_m(l)$, which represents the reflection coefficient at the receiving end of the network at $x = l$. According to Fig. 2, $\Gamma_m(x)$ may be defined as follows:

$$\Gamma_m(x) = \Gamma_m(x = l)e^{-2\gamma(l-x)} = \Gamma_m(x = l)e^{2\gamma x}e^{-2\gamma l} \quad (1)$$

For the ideal load that was imagined in Fig. 2, the reflection coefficient at $x = 0$ may also be defined as $\Gamma_m(x = 0) = \Gamma_m(x = l)e^{-2\gamma l}$. In general, according to electromagnetic properties, the coaxial probe problems such as those in Fig.2 are used to make the TEM mode analysis for propagation along the transmission line. Therefore, the transmission lines can be considered as a one port microwave network, in which the microwave power can enter or leave through only one transmission line. We assume that this one port network can be fully understood with a scattering parameter, $S_{11}$, by using the coaxial probe (see Fig.2). An ordinary text book presents a useful formula for these types of circuits, which is the following [9]:

$$S_{11} = \frac{\Gamma_m(e^{-2\gamma l} - 1)}{1 - \Gamma_m^2 e^{-2\gamma l}} \quad (2)$$

The equation (1) may be derived from the following formats too:

$$S_{11}(x = 0)\Gamma_m(l) = \frac{\Gamma_m^2(l)(e^{-2\gamma l} - 1)}{1 - \Gamma_m^2(l)e^{-2\gamma l}} = -1 + \frac{1 - \Gamma_m^2(l)}{1 - \Gamma_m^2(l)e^{-2\gamma l}} \quad (3)$$

$$S_{11}(x = 0)\Gamma_m(l) + 1 = \frac{1 - \Gamma_m(l)^2}{1 - \Gamma_m^2(l)e^{-2\gamma l}} \quad (4)$$

According to equation (3), one would need to obtain a reference plane with an ideal reflection, $\Gamma_m(l) \approx \pm 1$. This may be possible for the coaxial probe experiments. As for an ideal load case ($\varepsilon_m \cong 56 - j0$ in Fig.2), it plays the role of attaining a matched network. There are to be 50 ohms from x=0 to x=l, and the matched network would create a perfect transition, with no reflection between x=0 and x=l in the circuit. By using the reflection definition of equation (1) and the circuit structure in fig.2, one obtains the following:

$$\varepsilon_m = \left(\frac{Z_0}{Z_m(x = 0)Z_m(x = l)}\right)^2 \frac{1 + \tilde{\Gamma}_v}{1 - \tilde{\Gamma}_i} \quad (5)$$

where

$$\tilde{\Gamma}_v = \Gamma(l) + \Gamma(l)e^{-2\gamma l} + \Gamma(l)^2 e^{-2\gamma l}$$

$$\tilde{\Gamma}_i = \Gamma(l) + \Gamma(l)e^{-2\gamma l} - \Gamma(l)^2 e^{-2\gamma l} \quad (6)$$

To be an ideal load, the load ($\varepsilon_m$) has to be in quarter wave lengths, and

$$Z_m(x = 0) = Z_m(x = l) = Z_m = \frac{120\pi}{\sqrt{\varepsilon_m}} = 50 Ohm.$$

It is clear that the maximum reflection is now possible for the ideal load, and it can be expressed as $\Gamma_m(l) \approx \pm 1$. This is because $Z_0 = 120\pi$ ohms and $Z_m(l) = 50$ ohms, so one obtains the follows: $\Gamma_m(l) = [Z_0 - Z_m(l)][Z_0 + Z_m(l)]^{-1} \approx 1$. As such, equation (4) simplifies to the following:

$$S_{11}(x = 0)\Gamma_m(l) + 1 \approx 0 \quad (7)$$

By using the impedances relationships of the network (see Fig. 2) one may also write the following configurations as well:

$$\frac{Z_m(l)}{Z_{50}} = \frac{1 + \Gamma_m(l)}{1 - \Gamma_m(l)} = -\frac{1 - S_{11}(x = 0)}{1 + S_{11}(x = 0)} \quad (8)$$

$$\frac{Z_m(l)}{Z_{50}} = \frac{1}{\tanh(\gamma l)} \quad (9)$$

By combining (8) and (9), one can obtain the following relationships.

$$\left(\frac{Z_0}{Z_{50}}\right)^2 \frac{1}{\varepsilon_m} = -\left(\frac{1 - S_{11}(x = 0)}{1 + S_{11}(x = 0)}\right)\frac{1}{\tanh(\gamma l)} \quad (10)$$

$$\varepsilon_m = -\left(\frac{Z_0}{Z_{50}}\right)^2 \left(\frac{1 + S_{11}(x = 0)}{1 - S_{11}(x = 0)}\right)\tanh(\gamma l) \quad (11)$$

By now, it should also be convincing that the reflection coefficient is caused by the two way travel of propagating waves, so the propagation constants may be defined as follows:

$$\alpha_l = \alpha \ell = \frac{1}{2}\left(\frac{1}{8.68}\right)20\log_{10}|S_{11}| \quad (12)$$

$$\beta_l = \beta \ell = \frac{1}{2}\varphi(S_{11}) \quad (13)$$

This approach of the ideal load should be useful for the dielectric measurement of a real material media using a coaxial probe, since the measured $S_{11}$ can always be reasonable at $x=0$, which may include the circuit mismatches at their boundaries. As for the biological medium, the permittivity equation, equation (11), must be rescaled with regard to the permittivity of the known materials such as de-ionized or distilled water for this study. This is because the biological medium is similar to water. It is practical to use the distilled or de-ionized water as a reference material for the measurements of biological tissues. Research also shows that the traceable "known" complex permittivity is crucial for the reference liquids' purpose of measuring biological tissues [10, 11]. That also supports the rescaling of the permittivity equation (11) for water. The rescaling process is very simple:

$$\frac{\varepsilon^{ref}(\omega_0)}{\varepsilon^{ref}(\omega_0)}\varepsilon_m(\omega_0) = A(\omega_0) \text{ and } \varepsilon_{ref}(\omega) = \frac{\varepsilon_{ref}(\omega_0)}{\varepsilon_m(\omega_0)}A(\omega) ,$$

where A represents the right hand side of equation (11) for the frequency spectra. As such, one obtains the following:

$$\varepsilon_w(\omega) \cong -\left(\frac{\varepsilon_w^{ref}(\omega_0)}{\varepsilon_m}\right)\left(\frac{Z_0}{Z_{50}}\right)^2\left(\frac{1+S_{11}(x=0)}{1-S_{11}(x=0)}\right)\tanh(\gamma l) \quad (14)$$

As for the reference water measurements, the empirical rescaling factor was needed and was implemented as follows:

$$\frac{\varepsilon_w^{ref}(\omega_0)}{\varepsilon_m} = \left(\frac{80}{\varepsilon_m}\right)[1-0.01(1+j)f].$$ The empirical rescaling factor of frequency dependence was required in order to have better agreements with water data of earlier measurements published elsewhere. Perhaps, it implies that this method is still needed to consider neglected higher order modes when frequencies increase in the measurements. However, this kind of rescaling is sufficient for the measurements in this paper. As such, this rescaling factor is in accordance with the real permittivity of water for up to 18 GHz. As for the imaginary permittivity of water, it is in accordance up to 10 GHz, but becomes a little less reliable in the 10 GHz to 18 GHz range, compared to previous researchers. It may still be that the higher order modes are affected by the level of the signal at the higher frequency end. Once measurements for the referenced materials such as de-ionized and distilled water are established, it will be fairly simple to determine the dielectric properties of the biological tissues by repeating the method of equation (14). This can be seen below.

$$\varepsilon_t(\omega) \cong -\left(\frac{\varepsilon_w^{ref}(\omega_0)}{\varepsilon_m}\right)\left(\frac{\varepsilon_t^{S_{11}}(\omega)}{\varepsilon_w^{S_{11}}(\omega)}\right)\left(\frac{Z_0}{Z_{50}}\right)^2\left(\frac{1+S_{11}(x=0)}{1-S_{11}(x=0)}\right)\tanh(\gamma l)$$
$$(15)$$

It is obvious that the permittivity of the test material, $\varepsilon_t^{S_{11}}(\omega)$, is unknown, as well as the permittivity of reference water, $\varepsilon_w^{S_{11}}(\omega)$, without measuring the permittivity of those biological substances. However, it is possible to determine their ratios $\left(\frac{\varepsilon_t^{S_{11}}(\omega)}{\varepsilon_w^{S_{11}}(\omega)}\right)$ by using their phase measurements from calibrated reflections of $S_{11}$ at lower frequencies. This is done by considering the ratio of penetration distances of traveling waves in the different biological media. It is also possible to find the permittivity ratios from the propagation constants. In doing so, one may determine the reasonable ratios between penetration thicknesses that are supported by the Cole-Cole method. That process may be written as follows:

$$\gamma = \alpha + j\beta = j(\beta - j\alpha) = j\omega\sqrt{\mu_0\varepsilon_0(\varepsilon^r - j\varepsilon^i)} \quad (16)$$

By squaring both sides of equation (16), one can arrive at the following:

$$\frac{(\beta_1^2 - \alpha_1^2)l_1^2}{(\beta_2^2 - \alpha_2^2)l_2^2} = \frac{\varepsilon_1^r l_1^2}{\varepsilon_2^r l_2^2} \quad (17)$$

$$\frac{2\alpha_1\beta_1 l_1^2}{2\alpha_2\beta_2 l_2^2} = \frac{2\alpha_{l_1}\beta_{l_1}}{2\alpha_{l_2}\beta_{l_2}} = \frac{\omega^2\varepsilon_1^i l_1^2}{\omega^2\varepsilon_2^i l_2^2} \quad (18)$$

Here, $l_1$ and $l_2$ are penetration distance of wave propagation through the different medium. The $\alpha_1$, $\alpha_2$, $\beta_1$, and $\beta_2$ are the attenuation and propagation constants of propagating waves in respect to the different media. Also, the following notes were used in the calculation:

$\alpha_{l_1} = \alpha_1 l_1$, $\alpha_{l_2} = \alpha_2 l_2$, $\beta_{l_1} = \beta_1 l_1$, and $\beta_{l_2} = \beta_2 l_2$. As such, one obtains the ratios of penetration thicknesses as:

$$\frac{l_1}{l_2} = \left(\frac{\varepsilon_2^i}{\varepsilon_1^i}\right)^{\frac{1}{2}}\left(\frac{\alpha_{l_1}\beta_{l_1}}{\alpha_{l_2}\beta_{l_2}}\right)^{\frac{1}{2}} \quad (19)$$

The equation (17) is used to determine the initial and final permittivity of test materials with regards to the reference permittivity. It is based on the reasonable assumption that the penetration thickness would be infinite at both zero and infinite frequencies. That makes us eligible to detect the permittivity differences between the test and reference materials at the lossless region according to the Cole-Cole plot. The ratio of penetration thickness for the entire frequency range is calculated using equation (19), and is needed for the imaginary permittivity component of the test material frequency dependence from the Cole-Cole method as follows:

$$\varepsilon^i = imag \left( \varepsilon_\infty + \frac{\varepsilon_{00} - \varepsilon_\infty}{1 + j\dfrac{f}{f_0}} \right) \quad (20)$$

Here, $\varepsilon_{00}$ is the permittivity at zero frequency, $\varepsilon_\infty$ is the permittivity at infinite frequency, and $f_0$ is the relaxation frequency of medium.

### III. MEASUREMENTS OF WATER AND BIOLOGICAL SUBSTANCES

Usually, the coaxial probe measurement technique requires a simple calibration procedure, such as SOL. The standard open, short and match load (SOL) were used in this paper to remove potential reflections from the coaxial cable to the probe tip (see Fig.3). This is because of the potential reflection from coaxial cable and because its phase contributions have to be removed before starting the measurements. The phase removing procedure fully complies with Agilent's 8510C VNA built in SOL (short, open, and load) procedure.

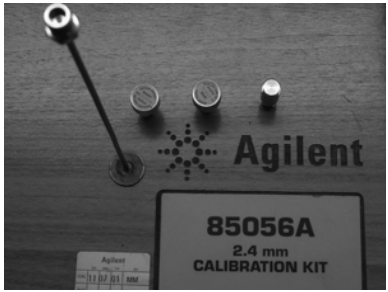

Fig. 3 The short, open, load standards and the coaxial probe were used in this measurement methodology

It implies that the 50 ohm match load that was implemented in the SOL calibration finds a plane that produces reflections that are typically lower than -50dB ($20\log_{10}|S_{11}| \le -50dB$). Afterwards, the match load is simply replaced by the 50 ohm coaxial probe (see Fig.3), and the coaxial probe replacement finds another plane with maximum reflections, i.e. $20\log_{10}|S_{11}| \approx 0dB$. At this point a plane with maximum reflection in regards to obtained magnitude of $S_{11}$ has been found. In the real time measurement process, the $S_{11}$ coefficient from the coaxial probe itself was recorded first as a reference. It was then used to normalize the $S_{11}$ magnitude and to find the $S_{11}$ phases from the material under test. To determine the normalized $S_{11}$ magnitude and its phases from the material under test, the simple algebra of the least squares method of polynomials was used for smoothing the data of the $S_{11}$ parameters as part of the calibration procedure. This simple reflection removal procedure provides an arrangement which is capable of returning the highest signal content from the coaxial probe tip attached to the test materials. In order to demonstrate the validity of this theory and the calibration procedure, the measurements of Methanol (HPLC grade, 99.9%) were presented, with the distilled water as the reference (see Fig. 4 and Fig. 5). The measured permittivity of methanol in Fig. 5 matches well with Agilent's measurements up to 10 GHz, but beyond 10 GHz the imaginary permittivity measurements achieved by this method are lower than the Agilent results [12]. Next are the presentations for the measurements of water and biological tissues. From these measurements in Fig.4, Fig.5, and Fig.6, there is a visible difference between the complex permittivity of deionized water and that of tap water, and these measurement differences are consistent for multiple trials. The measurements show that this technique is capable of differentiating the degree of the purity of the water (see Fig.6), and therefore can be considered a useful technique for detecting both normal and malignant substances in biological tissues. As for demonstrating the differentiation capability of this technique, the measured real and imaginary parts of complex dielectric permittivity were presented for commercial (available in grocery stores) beef and chicken materials using distilled water as a reference material. This new data is shown in Fig. 7 and Fig. 8.

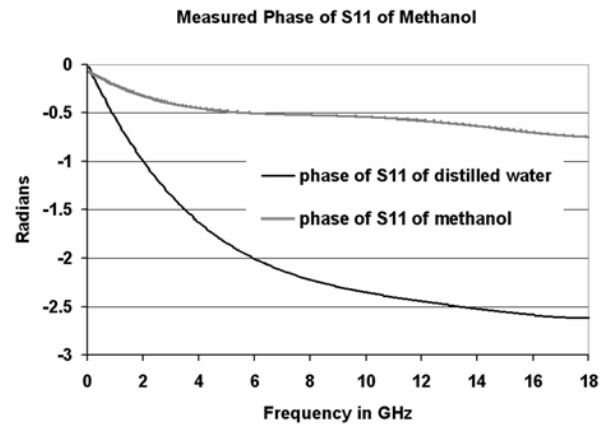

Fig. 4 $S_{11}$ phase measurements of distilled water and Methanol over the frequency range 0. 045- 18 GHz.

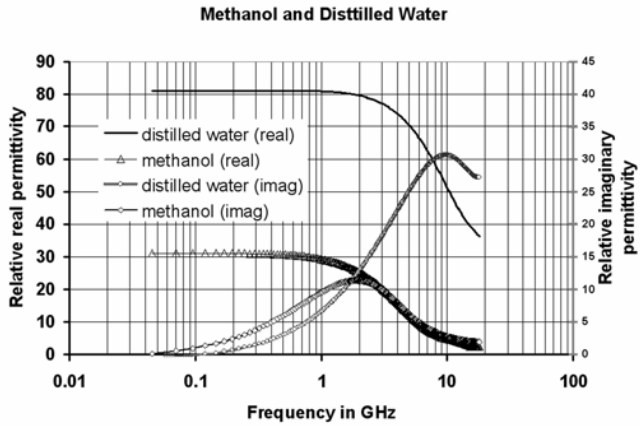

Fig. 5 Microwave permittivity spectra for the real and imaginary parts of complex dielectric permittivity of methanol and distilled water. The frequency is shown on a logarithmic scale.

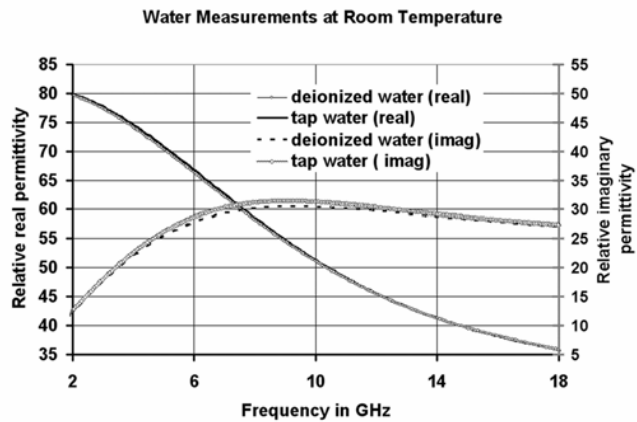

Fig. 6 A comparison of real and imaginary parts of complex permittivity spectra for deionized water and tap water. This technique reveals the difference between pure and tap water contents according to this measurement.

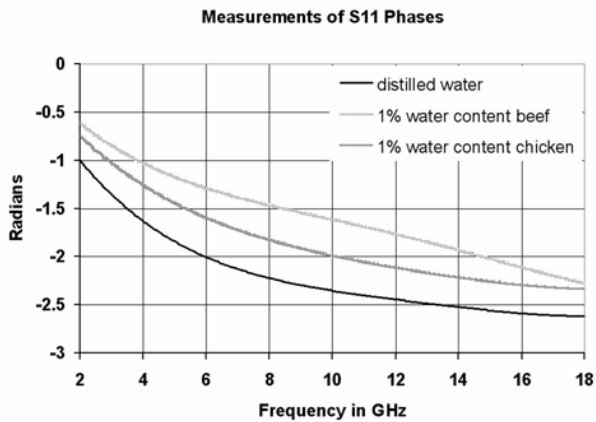

Fig. 7 the measured phase spectra for the scattering parameter $S_{11}$ for distilled water, commercial beef, and chicken (available in food stores). The typical retained water content for the beef and chicken is about one percent.

The tested beef and chicken meats were purchased from a usual food store and they were labeled as 1% retained water content, but not identified for the content of salt. Thus, unfortunately, the salt contents could not be presented in this paper. Fig. 7 compares the spectra for the phase of the scattering parameter $S_{11}$ for distilled water and commercially available beef and chicken. The experimental measurements of the above samples were carried out at room temperature in a microwave measurements lab. The reproducibility of the data is reasonable (see Fig.8). Following these tests, the measured complex permittivity of normal and malignant breast tissues were from the human body was presented in Fig.9 and Fig.10. The specimens from human body were collected from the Tufts New England Medical Center (Department of Pathology), and were formalin treated (10%). Specimens were then brought in to the microwave measurement laboratory for dielectric measurement purposes. The specimens have not been identified for race or age at this time. The measured results agree very well with the results that were reported by various other researchers (former experiments as mentioned in references) [16]. The results in this study were presented against the results and predictions of previous researchers, and in doing so; a useful table was created for comparison purposes (see Table 1). The table displays the dielectric properties of some of the materials measured in this experiment (see Fig. 8), in addition to references to previous experiments.

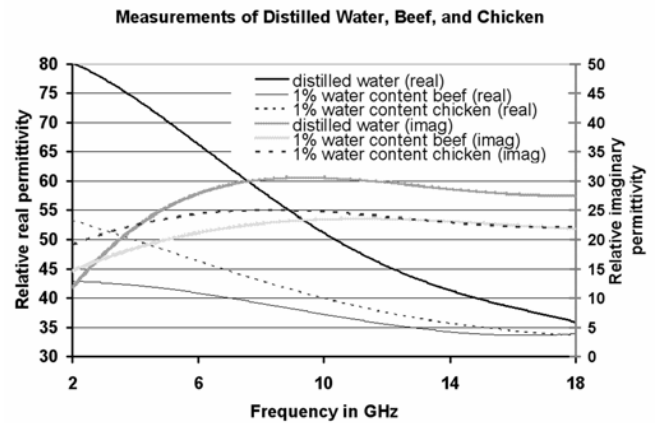

Fig. 8 Comparison of real and imaginary parts of complex dielectric permittivity spectra for distilled water, beef and chicken (available in grocery stores). Both beef and chicken contain about one percent water.

Findings include malignant tissue having more calcium content, making the real part of permittivity lower than that of water. Thus, one expects to see a higher imaginary part of permittivity for malignant tissues.

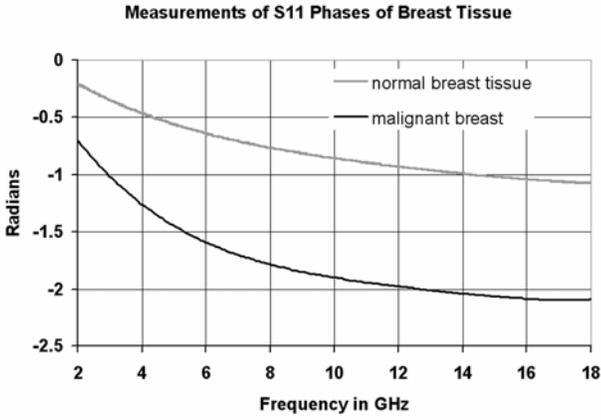

Fig. 9 the spectra showing $S_{11}$ phase measurements of 10% formalin treated normal and malignant breast tissues from humans. No information about the source is known yet.

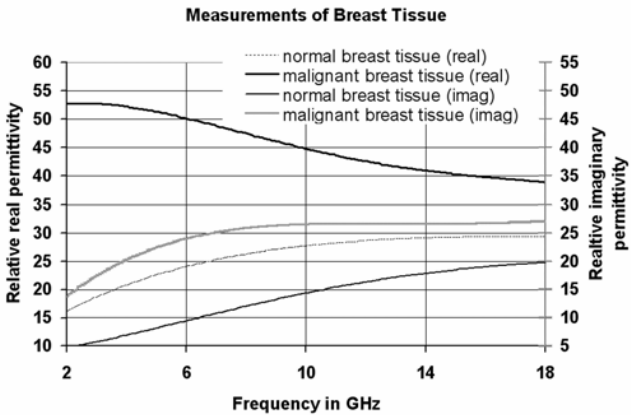

Fig. 10 Comparison of real and imaginary parts of complex dielectric permittivity spectra for 10 percent formalin treated normal and malignant breast tissue from human body.

Usually, the following mathematical curve fitting is used to apply for smoothing the data and errors analysis are employed for experimental measurements:

$$R = 1 - \frac{(y - y_{fit})^2}{(y - y_{mean})^2} \qquad (13)$$

Here, y stands for measured parameters, $y_{fit}$ being the fitted parameters from the least squares of polynomials, $y_{real}$ being the real part of the measured parameters, and $y_{mean}$ being the average value of the measured parameters. The value of R is always brought as close to one as possible, so that the best fit is achieved. A relative error analysis may be viewed in Fig.11. As for the errors analysis for other measurements, such as biological tissues, this study presents average absolute deviation (AAD) from the actual measurements, that is $AAD = \frac{1}{N} \sum_{i=1}^{N} |y_i - y_{mean}|$. The

$y_i$ stands for the measured data of each frequency spectrum. Fig.12 displays the AAD.

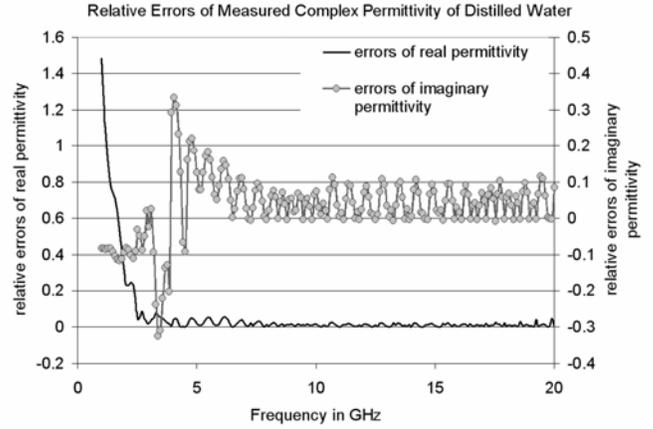

Fig.11 Relative error analysis of the complex permittivity of distilled water. The measured errors of relative real and relative imaginary parts were on a reasonable scale for supporting the reliability of this technique.

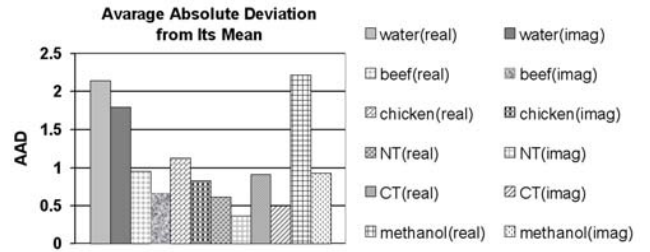

Fig.12 Average absolute deviations of the complex permittivity of six different substances which were measured by this technique

Table 1 Tabulated Permittivity and Conductivity Results

| Tissues | $\varepsilon_r'$ | $\varepsilon_r''$ | $\sigma_s$ | GHz | Ref |
|---|---|---|---|---|---|
| Beef | 43 | 13.5 | - | 2.45 | [12] |
|  | 47.7 | 13.4 | - | 2.8 | [14] |
|  | 42.7 | 16.5 | - | 2.8 | ours |
| Chicken | 52.3 | 17.7 | - | 2.4 | [12] |
|  | 37 | 5.0 | - | 12.0 | [12] |
|  | 37.6 | 23.8 | - | 12.0 | ours |
| Normal breast | 10~3 | - | 0.15~0.46 | 6.0 | [15] |
|  | 5 | 9.5 | - | 6.0 | ours |
|  | 24 |  |  |  |  |
| Malignant breast | 54 | - | 0.7 | 6.0 | [15] |
|  | 50.1 | 24 | - | 6.0 | ours |
| Fat%(0-30) | 48.4 | - | 0.7 | average | [16] |
| Fat%(85-100) | 4.7 | - | 0.036 | average | [16] |

It is important to differentiate errors by human, instrumentation, and system in a study with scientific measures. However, in this study, the above errors analysis is sufficient at this stage of the study. This is because the reported relative errors were already on a reasonable scale (within 5% fluctuations) for this report. See Fig.11 for distilled water and Fig.12 for average absolute deviations for each real and imaginary permittivity measurements by using this technique. Also, Fig.12 show the absolute average deviations, as well as displays that the liquid substances of water and methanol have larger average absolute deviation in comparison to the tissue substances from both the human body and cows and chickens. Logistically, the liquids have less interference problems compared to semi-solid substances such as tissues. As such, this source of error may not be mechanical from interference between coaxial probe tip and substance. Perhaps, system errors from calibration procedure may be reconsidered to eliminate the uncertainty region for its calibration standards. However, the porous media, such tissues, are usually composed of the materials of a solid matrix, a gaseous phase, and liquid water. Furthermore, the liquid water phase is sometimes subdivided into free water and bound water, which may be restricted in its mobility by specific structures, such as tissues. If so, the errors may be fundamental rather than caused by instrumentation. In either case, those kinds of errors analysis works should be reported in separate experiments. Perhaps that kind of study would help to differentiate the errors from different cell structures in tissues.

## IV. DISCUSSION AND CONCLUSION

As for the measured data for water, one should note that this technique is capable of showing the difference between de-ionized water and distilled water. The measurements show that a significant difference exists between tap water and de-ionized water in terms of the imaginary components of their permittivity. The complex permittivity measurements show that there are larger differences between beef and chicken from a food store. It is interesting that the white chicken meat has less fat content compare to the red beef meat that is consistent to the analysis of permittivity [17]. The measurements are in concurrence with previous studies of the permittivity of liquids, beef and chicken [13, 14, and 15]. These results certainly confirm the validity of this measurement technique. The normal and malignant breast tissue measurements also agree with previous studies [16, 17]. However, the normal breast tissue measurements showed some irregular behaviors at the lower frequencies of the spectrum. This phenomenon is attributed to the potential porosity of the normal breast tissue. The breast tissues from the human body were preserved in formalin liquid. The malignant breast tissue is a hard biological substance that would not absorb the formalin, but the normal breast tissue may have absorbed some amounts of formalin, creating certain porosity in the biological substance. This can be seen by the dispersion of the real permittivity (see Fig. 10) at lower frequencies in the spectrum. Also, it appears that the inference problem of this technique may need to be addressed in the near future. Finally, it is apparent that this technique will not have any conflicts with Agilent's standard probe technique. Alternatively, this method provides a new perspective for the microwave characterization of biological tissues and liquid substances.

### ACKNOWLEDGMENT

This research is supported by a contract from the US Army National Ground Intelligence Center.

# Permeability and Permittivity Measurement Technique by Propagation and Impedance of Microstriplines

*Abstract*—Numerous wireless communication devices such as integrated circuits and micro biochips are emerging in the 1 to 4 GHz frequency range. Thus it has become crucial to accurately determine effective permeability and permittivity of such oxide materials and devices in this spectral range. Traditional techniques such as waveguide and free space are good measurement techniques at higher frequencies. However, at lower frequencies it is extremely challenging to apply these methods. In this paper, we present a propagation and impedance technique for the microstripline. Three different substrates (Alumina, Silicon dioxide and Beryllium Oxide) are employed to design TRL sets for measurement purposes. The custom-designed TRL sets return losses up to -50 dB which is the same as a standard waveguide TRL calibration. Thin low-loss oxide samples were placed on the top side of a zero reference plane of an L microstripline. For the upper frequency band, we reported the permeability and permittivity measurements of the materials by waveguide without using any guess parameter. Here, we report effective permittivity and permeability measurements for oxides without using any guess parameter by microstripline. In the paper we present results of YIG, Nickel ferrite, and Glass from 1 to 4 GHz. Also, a nonlinear excitation of YIG is presented for demonstration purpose.

*Index Terms*—Permeability, permittivity, and microstripline TRL calibration.

## I. INTRODUCTION

Today's rapid developments in integrated circuits for various electronic, microwave and biomedical applications are based on thin film materials. The thin films are grown on various dielectric and magnetic substrates. They may be single layer or multilayer structured. For the bulk of available materials, the waveguide technique is suitable to determine their relative permeability and relative permittivity in air. The machined thin substrates have dies of various microstripline circuits on their surfaces. Since the relative permittivities of substrates are known, the impedance and effective relative permittivity of microstripline circuits may be configured by Wheeler equations [1] for circuit design purposes. Microstripline was widely used and recognized for microwave technology applications [2, 3] for magnetic materials, such as YIG, in the past few decades. The application of the microstripline technique and the TRL (thru-reflect-line) calibration method for microstriplines are not new. However, there is a new perspective proposed here for applying a direct measurement technique by using microstriplines to find the permittivity and permeability of thin films of single and multilayered materials. This technique allows for the detection of effective permittivity and permeability of oxides, including magnetic oxides at frequencies between 1 GHz to 4 GHz. The use for such an application was observed previously, when a microstripline technique for the simultaneous measurement of permittivity and permeability of materials was revealed [4]. However, this technique featured a full wave analysis for the microstripline, where the full wave propagating through the microstripline may not be in a TEM mode. Consequently, a numerical optimization process was used to find global minima in the squared error analysis, which may need more testing in order to confirm the validity of the algorithm. However, some papers considered the accuracy of the quasi TEM mode configuration for microstriplines in a propagation and impedance analysis [5, 6]. Based on quasi TEM mode assumptions, measurement techniques by microstripline were also presented elsewhere to measure permeability of ferrites, but these techniques require a microstripline fabrication on the material in question [7, 8, 9]. In order to make the microstripline technique more capable and flexible for measuring permeability and permittivity simultaneously, a new microstripline technique is presented in this paper. It is capable of simultaneously measuring the effective permeability and permittivity of thin materials without requiring a microstripline fabrication on either the thin oxide slab of the target or on films. In order to effectively determine the permeability and permittivity of the sample being tested, this paper articulates the valuable microstripline concepts and transmission line theory that have been published by others working in this area [10, 11]. Also in this paper is an attempt to see the nonlinear excitations from the YIG samples. Nonlinear excitations were studied and reviewed for YIG materials to reveal the fundamentals of nonlinear excitations of the YIG samples [12, 13, 14, and 15]. Evidently, this technique will also provide a reasonable platform for studies of nonlinear excitations in magnetic thin films as well.

## II. THEORY AND MEASUREMENT TECHNIQUE

Although numerous techniques, such as in-waveguide, cavity perturbation and free space are available for the permeability and permittivity measurements at higher microwave frequencies, the applicability of those techniques at lower frequencies is challenging due to the physical constraints of measurement equipment and instruments. The waveguide dimension becomes cumbersomely robust, thus requiring several pounds of materials for any reasonable measurement. However, at present there is a huge commercial interest for operating frequencies lying between 1 GHz and 4 GHz or even broader frequencies up to 10 GHz for various wireless communication applications [16, 17]. Thus, a reliable, cost- effective and fast measurement technique may be needed for the effective permeability and permittivity measurements of materials at lower frequency spectra. For this purpose, an obvious choice is a microstripline technique that does not need complex fabrication technology for the TRL standards. The custom designed standard TRL sets (see Fig.1 and Fig.2) of different dielectric substrates were fabricated at Microfab Incorporation (NH, USA). In this technique, the characteristic impedance of line standards is known prior to TRL calibration. This helps determine the characteristic impedance and propagation constant of the loaded material that has to be found by employing S-parameters of materials or devices being tested in a constructed network. The physical dimensions of the TRL sets are shown in Table.1. Since the requirement of loaded materials on the line standard are far less in volume compared to the substrates of the microstripline, a reliable reflection coefficient, $\Gamma$, is not expected from the loaded material on the microstriplines.

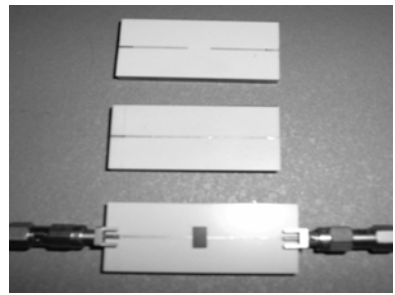

Fig. 4 A set of TRL BeO calibration substrates. The loaded magnetic material was placed on the microstripline for testing.

In general, for the cases of permeability and permittivity measurements using two port networks by $S_{11}$ and $S_{21}$, two different methods are available [10, 19]. The first is based on the concept of the existence of multiple reflections inside lossless media. Baker-Jarvis has presented the following, which is closely consistent with the well-known Weir equations in nature [18, 19]. That is equation set (1). This set of equations (1) has been in practice for decades in respect to the workable subjects.

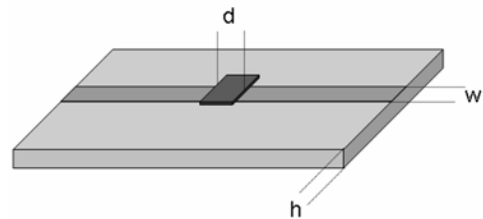

Fig. 5 A sketch showing geometrical dimensions of microstripline and material under test by a constructed network.

Table.1 Physical dimensions and electrical parameters for various microstripline substrates (1 mil =0.0254mm)

| Substrate | $\varepsilon_{eff}$ | $Z_0(\Omega)$ | w(mil) | h(mil) | (L-T) (mm) |
|---|---|---|---|---|---|
| SiO$_2$ | 3.07 | 50 | 46.6 | 25 | 5.95 |
| BeO | 4.58 | 50 | 29.9 | 25 | 3.96 |
| Alumina | 6.56 | 50 | 20.1 | 25 | 2.75 |

$$S_{11} = \frac{\Gamma_i(1 - T_i^2)}{1 - \Gamma_i^2 T_i^2}$$

$$S_{21} = \frac{T_i(1 - \Gamma_i^2)}{1 - \Gamma_i^2 T_i^2}$$

$$T_i = e^{-\gamma l} = \frac{S_{21}}{1 - \Gamma_i S_{11}} \qquad (1)$$

$$\Gamma_i = \frac{Z - Z_0}{Z + Z_0}$$

On the other hand, based on the transmission line theory, Vittoria has presented a new set of equations, which is as follows [10].

$$S_{11} = \frac{\Gamma_e(T_e^2 - 1)}{1 - \Gamma_e^2 T_e^2}$$

$$S_{21} = e^{-\gamma l}(1 + S_{11}\Gamma_e)$$

$$T_e = e^{-\gamma l} = \frac{S_{21}}{1 + \Gamma_e S_{11}} \qquad (2)$$

$$\Gamma_e = \frac{Z_0 - Z}{Z_0 + Z}$$

The question now is how far the equations from set (1) to set (2) are. Based on the definitions of their first reflections, it seems that $\Gamma_i = -\Gamma_e$. From this equality, the equation set (1) and (2) should be the same in nature. However, the problem is not so simple, since one must also consider where the reflections are coming from. Upon doing so, according to the definitions of first reflections ($\Gamma_i$ and $\Gamma_e$), one would correlate them and end with the relationship $\Gamma_i = e^{-2\gamma l}\Gamma_e$. This implies that if the network is lossless and the used transmission line distance between thru and line equals a quarter wavelength, then the James and Vittoria equations hold and are equivalent in nature. For the low frequency measurements using the microstripline in this study, the equation set (2) was only partially used, the obvious reason bring that the loaded materials on the transmission lines of TRL calibration standards used in this study were not capable of generating reliable multiple reflections for the quarter wavelength differences between the thru and line of the microstripline for this frequency spectra. The Baker-Jarvis equation for the waveguide measurement technique was used in that study [20], because the transmission line differences between the thru and line of TRL standards of waveguides were on the order of quarter wavelengths, and the loaded materials inside the waveguide are able to generate reliable reflections [20]. In summary, there is the option to choose between equation sets (1) and (2) based on which problems are less pertinent. As for the perfectly calibrated matched transmission line, the relationship between the transmission coefficients would be $T = S_{21}$. However, the target material loaded transmission line is capable of generating a negligible amount of reflection from the scattering coefficient $S_{11}$, so the transmission coefficient through the material needs to be normalized to its loss of $|S_{21}|^2 = 1 - |S_{11}|^2$. Bearing this in mind, it is noted that the reasoning behind the normalization has to be supported by reasonable methods. According to the set of equations (2) we have the option to deploy its transmission as follows: $T = e^{-\gamma l} = S_{21}(1 + S_{11}\Gamma_L)^{-1}$. Due to the nature of the problem being addressed in this paper, reliable return reflection from the samples that were loaded on the top side of the transmission lines was unattainable. Also ignored were the arbitrary phases of small reflection errors of very small magnitude of the loaded sample in the calculations. Allow the normalization factor be defined as follows:

$$|T|^2 = TT^* = S_{21}S_{21}{}^* \left(1 + |S_{11}\Gamma_L|e^{-j\psi}\right)^{-1}\left(1 + |S_{11}\Gamma_L|e^{j\psi}\right)^{-1} \qquad (3)$$

Manipulating the above equation with simple algebra produces the following:

$$|T|^2 = |S_{21}|^2\left(1 + |S_{11}\Gamma_L|^2 + 2|S_{11}\Gamma_L|\cos(\psi)\right)^{-1} \qquad (4)$$

For equation (4) shown above, the higher order magnitude of reflections can be dropped due to negligible magnitudes from reflection. Also, one would assume $\psi \to 2m\pi$, (m= 0, 1, 2, etc) which allows for the removal of arbitrary phases of $S_{11}$ and $\Gamma_L$, in order to reach the following:

$$|T|^2 \cong |S_{21}|^2(1 + |S_{11}|^2)^{-1} \cong (1 - |S_{11}|^2)|S_{21}|^2 \qquad (5)$$

Now it is easy to obtain the magnitude of the transmission coefficient:

$$|T| = \sqrt{(1 - |S_{11}|^2)} \cdot |S_{21}| \qquad (6)$$

Now the transmission coefficient of this microstripline may be determined as follows:

$$T = e^{-\gamma l} = e^{-(\alpha + j\beta)d} = \left(\sqrt{1 - |S_{11}|^2}\right) \times S_{21} \qquad (7)$$

One can then extract the propagation constant in loaded material through the transmission coefficient as follows:

$$\gamma = \frac{1}{d}\left(\ln\left(\frac{1}{|T|}\right) + j(2\pi n - \varphi_T)\right) \qquad (8)$$

According to the equivalent circuit (see Fig. 3), the $A_{11}$ element of A-matrix may be represented by the elements of the S-matrix.

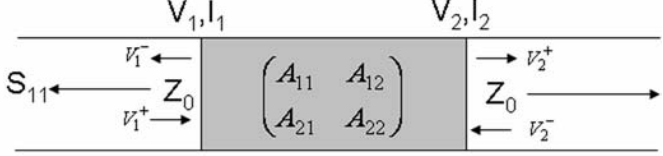

Fig. 6 Equivalent circuit diagram of microstripline. The $A_{11}$, $A_{12}$, $A_{21}$ and $A_{22}$ are elements of the A- matrix. $Z_0$ is the characteristic impedance, $V_1$, $V_2$, $I_1$ and $I_2$ represents voltages and currents

Now the relation between the elements of the A matrix and the S-matrix are presented, in accordance with the constructed network in Fig.3. The circuit may be represented as having a single voltage.

$$A_{11} = \frac{V_1}{V_2}\bigg|_{I_2=0} = \frac{1}{2}\frac{\left(\frac{V_1^+ - V_1^-\frac{V_1^-}{V_1^+} + V_2^+\frac{V_2^+}{V_1^+}\right)}{V_2^+} = \frac{1}{2}\left(\frac{1-S_{11}^2}{S_{21}} + S_{21}\right)$$

$$\qquad (9)$$

$$A_{21} = \frac{I_1}{V_2}\bigg|_{I_2=0} = \frac{I_1}{V_1}\frac{V_1}{V_2}\bigg|_{I_2=0} = \frac{2}{Z_{11}}\frac{V_1}{V_2}\bigg|_{I_2=0} \qquad (9a)$$

The currents in the networks were considered to be equal and in opposite directions, making the total current to be zero in the network, it is because of load oxides are no grounded in the circuit. According to the relationship between Z and the S matrix of the reciprocal network, one would have the relationship of:

$$Z = (I + S)(I - S)^{-1}$$

$$Z_{11} = Z_0\frac{(1+S_{11})(1-S_{22}) + S_{12}S_{21}}{(1-S_{11})(1-S_{22}) - S_{12}S_{21}} \qquad (9b)$$

The concentration of this paper was on the measurements of isotropic media that can stay in a reciprocal state in the network, meaning $S_{11} = S_{22}$ and $S_{12} = S_{21}$. By the combination of (9), (9a), and (9b), $A_{21}$ may be derived as follows:

$$A_{21} = \frac{1}{Z_0}\left(\frac{(1-S_{11})^2}{S_{21}} - S_{21}\right) \qquad (9d)$$

It is obvious that it is convenient for us to use an A matrix rather than an S-matrix alone to determine the permeability and permittivity from the microstripline technique. We employed a reasonable method for correlating the two matrices based on reciprocal network concepts that were presented by former investigators [6, 7, 8, 9, 10, and 11]. The relationship of the A-matrix and the S-matrix yields the result $A_{11} = A_{22}$, for example. As for the $A_{12}$, we need to solve the determinant of the A matrix, $\det(A) = 1 = A_{11}A_{22} - A_{12}A_{21}$ which produces a relationship of $A_{12} - A_{21}\frac{Z_0^2}{4} = \frac{S_{11}}{S_{21}}Z_0$. It is now possible to obtain the full elements of the A matrix with the following:

$$A = \begin{pmatrix} \cosh(\gamma d) & Z_L\sinh(\gamma d) \\ \frac{1}{Z_L}\sinh(\gamma d) & \cosh(\gamma d) \end{pmatrix}$$

$$= \frac{1}{2}\begin{pmatrix} \frac{1-S_{11}^2}{S_{21}} + S_{21} & \frac{Z_0}{2}\left(\frac{(1+S_{11})^2}{S_{21}} - S_{21}\right) \\ \frac{2}{Z_0}\left(\frac{(1-S_{11})^2}{S_{21}} - S_{21}\right) & \frac{1-S_{11}^2}{S_{21}} + S_{21} \end{pmatrix}$$

$$\qquad (10)$$

The normalized characteristic impedance of loaded materials may then be written as follows:

$$Z_L^n = \eta = \frac{\sinh(\gamma d)}{\left(\frac{(1-S_{11})^2}{S_{21}} - S_{21}\right)} \qquad (11)$$

Here, $Z_L$ is the load impedance on the microstripline, $Z_L^n$ is the normalized impedance on the microstripline of target materials, and $\eta$ is the medium impedance, in ohms. Based on a quasi TEM mode assumption of the microstripline on the substrate, a derivation of the relative permittivity and permeability of loaded materials to the L standards of the substrate may be done:

$$\varepsilon_{eff}^r = -j\frac{c}{f}\frac{\gamma}{2\pi}\frac{1}{Z_L^n}$$

$$\mu_{eff}^r = -j\frac{c}{f}\frac{\gamma}{2\pi}Z_L^n$$

$$\qquad (12)$$

Here, $\gamma_0 = j\frac{2\pi}{\lambda_0} = j\frac{2\pi f}{c}$, and c is the speed of light. The impedance of microstriplines fabricated on substrates is known, so it is possible to derive the relative permeability and relative permittivity of oxides by using (12).

### III. EXPERIMENTAL MEASUREMENTS

In this measurement process, the widths of the loaded materials are shorter than the line difference between L and T of TRL sets from different substrates. The samples are placed on the center of calibrated L lines. Locating the sample on the exact reference plane side of $S_{11}$ is always crucial for obtaining reliable phases of reflections, which help to derive accurate data for permittivity and permeability. For the measurements on $SiO_2$ and BeO substrates, the loaded material dimensions are width = 3.5mm, height = 0.5mm and baseline = 7mm. For the measurements on Alumina substrates, the loaded material dimensions are width = 1.26mm, height = 0.5mm and baseline = 2.52mm. In the final data extrapolation, the propagation contributions from the L lines ($\delta_l = d - \Delta(L - T)$) of the substrates were subtracted from the total propagation constant. They were recorded using a vector network analyzer. Next, the simultaneous permeability and permittivity measurements of thin ferrite disks and glass on the BeO substrate were done by this microstripline technique (see Fig.4, 5, and 6). The simple table below (Table 2) compares the data for real permittivity by microstriplines and by the in-waveguides of YIG, Nickel ferrite, and Glass on BeO substrate at 4 GHz for comparison purposes.

Table.2 A comparison of permittivity data at 4GHz for YIG, nickel ferrite and glass measured by two methods

| Materials | $\varepsilon_{eff}^{microstripline}$ | $\varepsilon_{relative}^{waveguide}$ |
|---|---|---|
| YIG | 6.8 | 13.0 |
| Nickel ferrite | 6.8 | 12.0 |
| Corning® 1737 glass | 5.5 | 6.0 |

It was already noted that the measurement figures from figs 4 to 6 show very smooth data for permeability and permittivity. Consequently, the measurement data show some unphysical parts for both permittivity and permeability. Also, the figures show some frequency dependent increases for the real part of the permittivity. Let it be mentioned that the equation (7) was based on a bold approximation of the magnitude of $S_{11}$ being very small. However, the reflection from any sample is increased when the frequency increases. As such, the approximation appears bold, but when the frequency increases, perhaps this will cause some frequency dependent phenomena for the permittivity in the figures above. Conversely, a method of least squares of polynomial for presenting smooth data on these figures has been applied. In general, the following mathematical curve fitting for smoothing the data is deployed for the experimental measurements.

$$R = 1 - \frac{(y - y_{fit})^2}{(y - y_{mean})^2} \text{ and } err = \frac{\Delta y}{y_{real}} = \frac{y - y_{mean}}{y_{real}} \quad (13)$$

Here, y stands for measured parameters, $y_{fit}$ stands for fitted parameters from the least squares of polynomials, $y_{real}$ is the real part of the measured parameters, and $y_{mean}$ stands for the average value of the measured parameters. In order to achieve the best fit, the value of R is made as close to one as possible. However, fitting that is too good creates some data for the imaginary components of the permeability and permittivity. Therefore, errors within 5% would be acceptable for our measurements. We attribute these 5% errors to system errors, instrumentation errors, and human errors, but the quantified differentiations of error sources will not be analyzed at this time. The 5% errors stand for compromising between best curve fitting and the errors that are allowed due to indications of unphysical parts measurements of the permeability and permittivity. This kind of curve fitting may also applicable in finding direct errors from the scattering parameters of $S_{11}$ and $S_{21}$ if it is necessary to do so. When using this curve fitting, the permeability and permittivity measurement figures show little unphysical ranges, and they are within the acceptable spectra that were presented in figs.7 and 8. Although this technique works very well for the simultaneous permeability and permittivity measurements of several materials on BeO substrate, the measurements of permittivity on the Alumina and $SiO_2$ substrates were not successful with this technique. In contrast, the complex permeability of the materials is detectable by using this microstripline on Alumina, BeO and SiO2 substrates. For a demonstration, nonlinear excitations obtained from YIG by using this microstripline technique on BeO substrate are presented in fig.9.

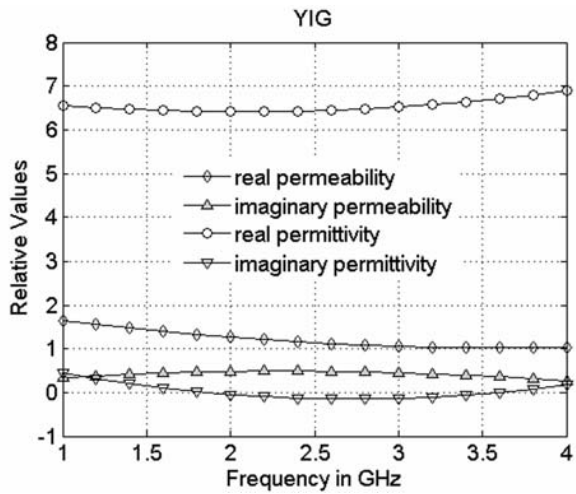

Fig. 4 Complex magnetic permeability and dielectric permittivity measurements of YIG on BeO substrate

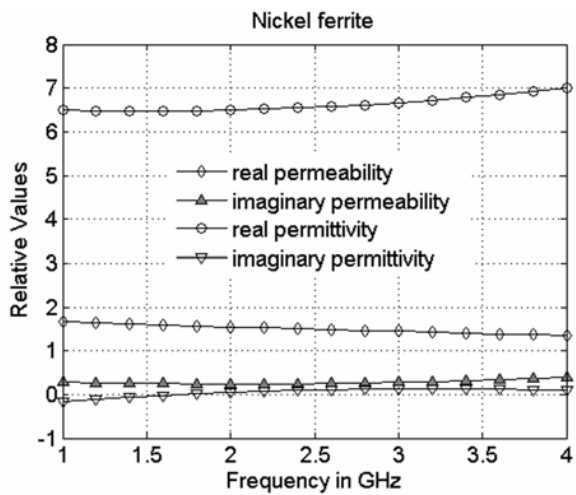

Fig. 5 Complex magnetic permeability and dielectric permittivity measurements of nickel ferrite on BeO substrate

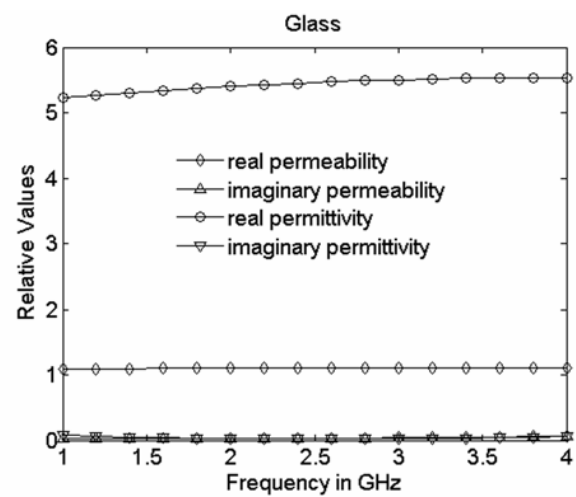

Fig. 6 Complex magnetic permeability and dielectric permittivity measurements of glass (Corning® 1737) on BeO substrate

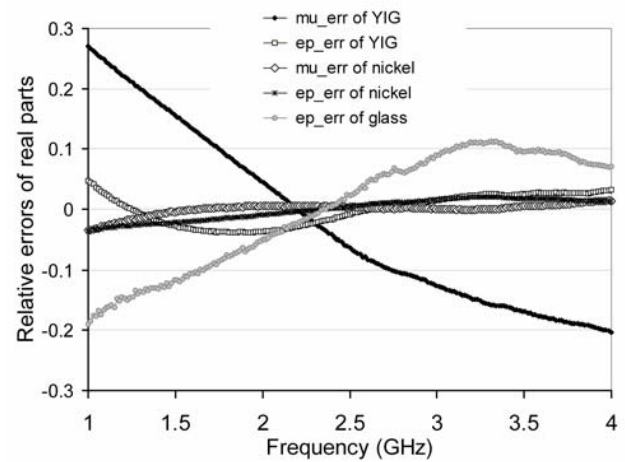

Fig.7 Relative errors of real parts of measured permeability and permittivity, where mu represents real permeability, ep represents real permittivity and err represents errors. YIG (yttrium iron garnet), nickel is nickel ferrite, and glass is corning glass 1735.

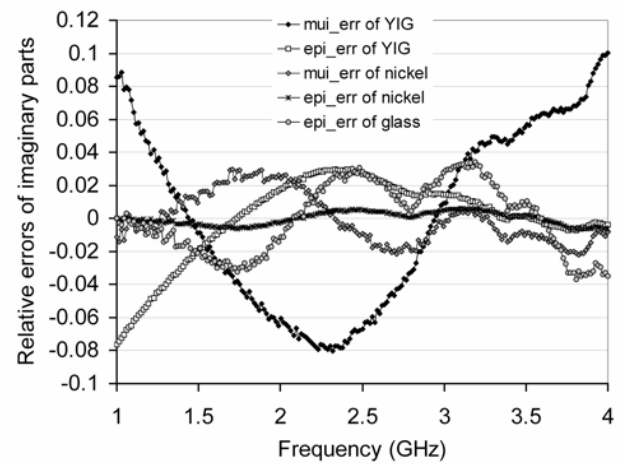

Fig. 8 Relative errors of imaginary parts of measured permeability and permittivity, where mui represents imaginary permeability, epi represents imaginary permittivity and err represents errors. The YIG (yttrium iron garnet), nickel is nickel ferrite, and glass is corning glass 1737.

As for the nonlinear excitation analysis of YIG, numerous excellent studies have already been reported in various journals for specific cases of YIG microwave measurements in the past several decades. For example, some of the nonlinear analysis and studies for YIG materials were reported to understand the nonlinear dynamical behaviors of YIG [12, 13, 14, and 15]. In this study, a YIG sample was simply placed on the microstripline and it was subjected to non-uniform magnetic field. To see the nonlinear excitations from YIG, the excited nonlinearity was correspondingly attributed to excited MSM cases (magneto static modes), according to the findings of former investigators. It is clear that the nonlinear phenomenon of YIG has been known for five decades, so a detailed discussion on the nonlinear phenomenon will not be presented at this time. It is beyond the scope of this paper, but it would appear that the nonlinear excitations of YIG might also be observed by using this technique without further complicated experimental setup. Perhaps it will provide a new scope to study the nonlinear excitations of YIG since the YIG has had a vital role in various technological applications for decades.

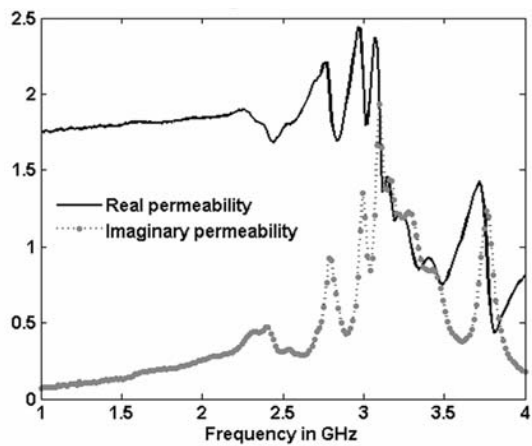

Fig. 9 A comparison of spectra showing the real and imaginary parts of magnetic permeability of YIG specimen. A YIG rectangular disk was placed in non-uniform fringe external fields of 500 Oe

## IV. DISCUSSION AND CONCLUSION

For the higher permittivity substrate such as Alumina, the propagation wave through the microstripline may not be a quasi TEM mode. As for the loaded magnetic specimen, perhaps, one may have to take into account its demagnetization effect on wave propagation. Since the measurements showed that a microstripline on Alumina substrate severely weakens permittivity measurements of the oxides at lower frequencies spectra, the quasi TEM mode approximation for microstripline on the Alumina substrate may not be true in this case. The microstripline technique for $SiO_2$ showed that higher imaginary components for complex permittivity of the materials, which we attribute to the conductive carriers in $SiO_2$. However, in this design, the fabricated microstripline on BeO fitted the quasi TEM mode approximation very well, which can be seen in the presented data on the measured effective complex permeability and permittivity of YIG, nickel ferrite, and glass material. The permittivity and permeability measurements are slightly different between the waveguide technique and the microstripline technique on dielectric substrates [20]. In an inside waveguide measurement technique, the materials were positioned in air, so that the measured relative permittivity and relative permeability were in relation to air. However, with the microstripline measurement technique, the measured permittivity and permeability were relative to the substrate and air. Thus, the presented complex permeability and permittivity can be understood as effective permeability and effective permittivity due to the averaging effect between the load, dielectric substrate and air.

## ACKNOWLEDGMENTS

This research is supported by a contract from the US Army National Ground and Intelligence Center.

# Full Band Microwave Isolator of Rectangular Waveguide Having Periodic Metal Strips and Ferrites

*Abstract*— **This study presents a full band microwave isolator in the X-band spectrum that features ferrite samples coated in uniformly spaced strips of metal wire. The objective is to use the wire strips to obtain effective negative permittivity levels at the desired frequencies within the spectrum, as well as to create negative permeability with the ferrite-wire configuration. In practice, we were capable of controlling the wire-covered ferrite samples by using an external field as small as 100 mT. The controllable permeability of ferrites in this experiment allowed us to create unidirectional wave propagation over the entire X-band frequency spectrum. This nonreciprocal circuit is a challenge to the traditional methods of defining refractive index, permittivity and permeability from the S-parameters of a vector network analyzer. We therefore propose a novel method to define the refractive index for nonreciprocal circuits.**

*Index Terms* — *metal wires, ferrites, negative permeability, negative permittivity, refractive index and metamaterials*

## I. INTRODUCTION

The microwave isolator has been in practice for several decades. The use of ferrites in waveguides for isolator application is now a well-established concept in the field of microwave technology. Although this type of study has a long history with regard to its known practical applications and theoretical completeness, we plan to construct a full band microwave isolator in the X-band spectrum using the wire-coated ferrite samples. The resonance loss properties of ferrites are used to design one-way transmission lines. These lines have a large percentage of energy propagation absorption by the ferrite in one direction, while permitting nearly lossless transmission in the opposite direction. The real consequence of such results depends on the reaction to large external magnetic fields while operating in the high frequency region of the spectrum. Although some hexaferrites are effective in reducing the external field influence at higher frequencies, the production of high-quality hexaferrites for general microwave spectra often involves extra costs and a comprehensive knowledge of hexaferrite crystal structures. By using high-quality yet less costly ferrites, including YIG and Nickel, as well as applying a weaker external magnetic field to the microwave isolators in the waveguide, we present a method for deriving negative permittivity from lined metal wire arrays on the microwave isolator. The basic notions of negative permittivity were gathered from Pendry's notable findings [1]. The ferrites used were YIG and Nickel and the metal wire was a thin aluminum tape. The metal wires, which were 2mm wide, were taped along the width of the rectangular ferrite pieces with 0.5 mm between each strip. The wrapped ferrite disks were then loaded into an X-band rectangular waveguide. The S-parameter measurements showed that the wrapped ferrite disks were fully capable of annihilating the $S_{21}$ transmission for the full X-band spectrum; that is to say that the achieved annihilation was up to 4 GHz bandwidth. We attribute this phenomenon to effective negative permittivity from the periodic metal wire array. During the application of a 100 mT external magnetic field that was perpendicular to the RF magnetic field of the $TE_{10}$ mode in the rectangular waveguide, the annihilated $S_{21}$ transmission was dramatically recreated for the full X-band spectrum. This phenomenon is due to the effective negative permeability of the ferrites. Also observed in this experiment was a unidirectional propagating wave mode. The opposite direction of propagation became evanescent with the application of a 100 mT external field. These results are most likely due to the left-handed properties of metamaterials [2, 3]. Thus, we were able to generate broadband effective negative permittivity using metal wires, and effective negative permeability by using a 100mT external magnetic field perpendicular to the RF magnetic field of the $TE_{10}$ mode in the rectangular waveguide. The successful microwave isolator was used in an X-band waveguide and its isolation frequency band was 4 GHz, ranging from 8 GHz to 12 GHz. The isolation stretched as wide as 40 dB. In this paper, the negative permittivity and negative permeability generated by the microwave isolator in the X-band are presented with the scattering parameters measured by a vector network analyzer (the Agilent 8510C vector network analyzer). The microwave isolator presented in this paper is used to apply the concepts of negative permittivity and permeability for practical microwave device applications. As for the negative refractive index creation with metal wire and with ferrite materials, it may be useful for device miniaturization [3]. Although Nicolson-Ross and Weir [4] and Baker-Jarvis [5] presented ways of defining the refractive index, permeability, and permittivity in a waveguide, their methods require initial estimated parameters and integer values for phases of propagating waves in normal microwave materials. It is often hard to be accurate with those estimations when using a metamaterial. As for determining metamaterial permittivity and permeability, a method has been presented by Smith, Schultz and Markos [2], who proposed that the constructed network or circuits should be in reciprocal states in order to do so. The circuit featured in this paper used neither normal materials nor remained in a reciprocal state. In the past, we presented a T/R method [6] for normal materials that does not need initial parameter estimations, and based on that method [6], we present a novel method for the special case of nonreciprocal circuits with metamaterials. Consequently, the method presented in this paper is successful in determining the permeability and permittivity of both normal materials and of reciprocal circuits from metamaterials.

## II. MEASURED S-PARMETERS OF YIG AND NICKEL FERRITE IN A WAVEGUIDE

In the experiment, the wires were made with rectangular aluminum foil strips. The dimensions of the wires are such that for w (width), L (length), and t (thickness), they satisfy the following condition: L > w >> t. The media is the lined wire array taped onto the ferrite disk and placed inside the waveguide. There were nine 2mm wide aluminum foil strips attached to the ferrite disks; with 0.5mm gaps separating them from one another in order to create effective negative permittivity (see Fig.1). Since the desired effective negative permittivity came from good conductors such as aluminum foil wires, the corresponding effective negative permeability was achieved by applying different external magnetic fields to the ferrite disks. In this experiment, the external field $H_0$ was applied along the RF electric field of the $TE_{10}$ mode in the rectangular waveguide. Thus, the $H_0$ field is parallel to the wires on the ferrite disk.

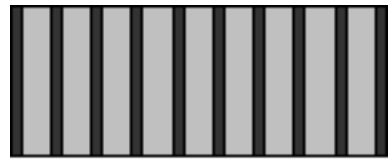

Fig. 7 Ferrite disk taped with aluminum wire strips. Aluminum wire width is 2 mm, the gap between wires is 0.5 mm, and the length of the disk is 2.3 cm.

The scattering parameters from the wire-coated ferrite samples in the waveguide may be viewed in Figures 2 and 3.

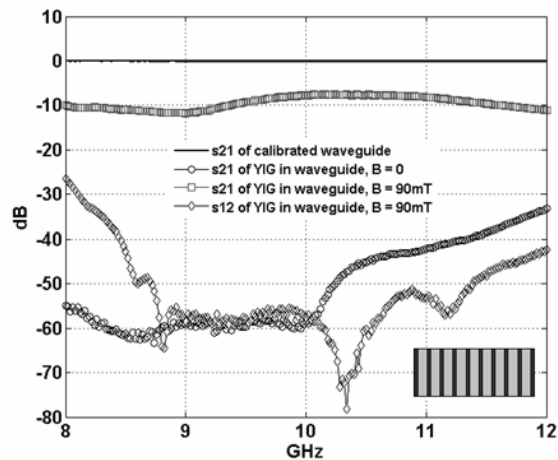

Fig.2 Measured scattering parameters of $S_{21}$ and $S_{12}$, in dB, from 8 to 12 GHz for aluminum wires taped on YIG with and without an externally applied magnetic field (90mT). Aluminum wire width is 2 mm, the gap between the wires is 0.5 mm, and length of the YIG disk is 2.3 cm.

For the case of external field, $H_0 = 0$ Oe, the $S_{21}$ transmission data shows that the transmission drops below -50dB (see Fig.2 and Fig.3). This indicates that the metal wires achieved negative permittivity and annihilated the transmission agent of $S_{21}$. While applying the external fields of 90 mT and 100mT to the wire-coated ferrite samples, the transmission agent $S_{21}$ was created back in the waveguide. This is simply because the negative permeability plays a role in changing the attenuated wave to the propagated wave again in the waveguide. Moreover, the $S_{12}$ gets a small advantage from the applied external fields. It should be noted that the right and left-handed permeability depends on the magnetic polarization by the external field, $H_0$, and its spin is caused by RF driving forces to the ferrite disks. Although we have 10 dB insertion losses, we have reached a very broad band of isolation between $S_{21}$ and $S_{12}$. The insertion loss can either be amplified by microwave amplifiers or improved upon using simulation designs. As was noted earlier, none of the available techniques are capable of determining the refractive index, permeability and permittivity of this kind of nonreciprocal circuit. It is now necessary to present a technique that is capable of deriving negative permittivity and permeability for such nonreciprocal circuits.

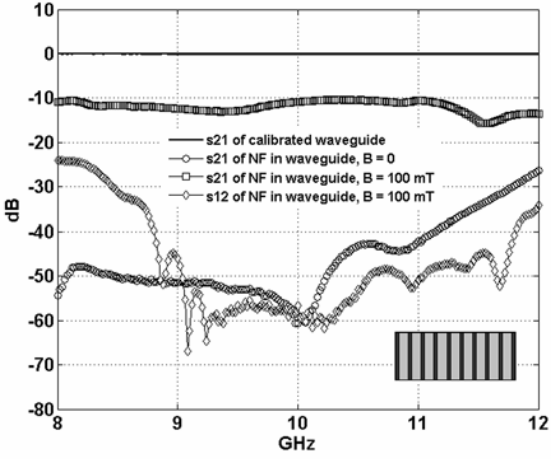

Fig.3 Measured scattering parameters of $S_{21}$ and $S_{12}$, in dB, from 8 to 12 GHz for aluminum wires taped on nickel ferrite with and without an externally applied magnetic field (100mT). Aluminum wire width is 2 mm, the gap between the wires is 0.5 mm, and length of the nickel ferrite disk is 2.3 cm.

### III. CAUCHY-RIEMANN EQUATIONS FOR THE PROPOGATION CONSTANT AND ITS APPLICATION TO THE REFRACTIVE INDEX DETERMINATION

The phase shift and attenuation from the transmission line of the waveguide [6] can be written as follows:

$$\Psi = j\gamma_0(n-jk)l = \varphi_0 k + j\varphi_0 n = \varphi_L + j\varphi_P \quad (1)$$

Applied to equation (1), the Cauchy-Riemann equation may be used as follows.

$$\frac{\partial \varphi_L}{\partial k} = \frac{\partial \varphi_P}{\partial n}$$
$$\frac{\partial \varphi_P}{\partial k} = -\frac{\partial \varphi_L}{\partial n} \quad (2)$$

Based on equations (1) and (2), we obtain the following:

$$n = \left(\frac{\partial \varphi_0}{\partial \omega}\right)^{-1}\left[\frac{\partial \varphi_P}{\partial \omega} - \varphi_0\frac{\partial n}{\partial k}\frac{\partial k}{\partial \omega}\right] \quad (3)$$

$$\frac{\partial k}{\partial \omega} = \frac{1}{\varphi_0}\left[\frac{\partial \varphi_L}{\partial \omega} - k\frac{\partial \varphi_0}{\partial \omega}\right] \quad (4)$$

$$\frac{\partial n}{\partial k} = \frac{\partial \varphi_P}{\partial \omega}\left(\frac{\partial \varphi_L}{\partial \omega}\right)^{-1} \quad (5)$$

By using equations (3), (4), and (5), one can be easily arrive at the following:

$$n = k\frac{\partial \varphi_P}{\partial \omega}\left(\frac{\partial \varphi_L}{\partial \omega}\right)^{-1} \quad (6)$$

In order to validate this measurement methodology, the normal material phenyloxide was tested first (see Fig.4).

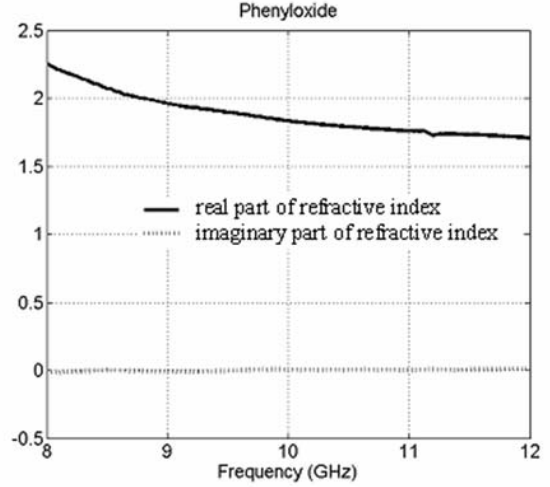

Fig. 4 Measurement of complex refractive index of normal material phenyloxide in waveguide using Cauchy-Riemann implementation.

As for the refractive index from the wire-coated nickel ferrite (see Fig.5), the measurement showed a negative refractive index up to 8.7 GHz.

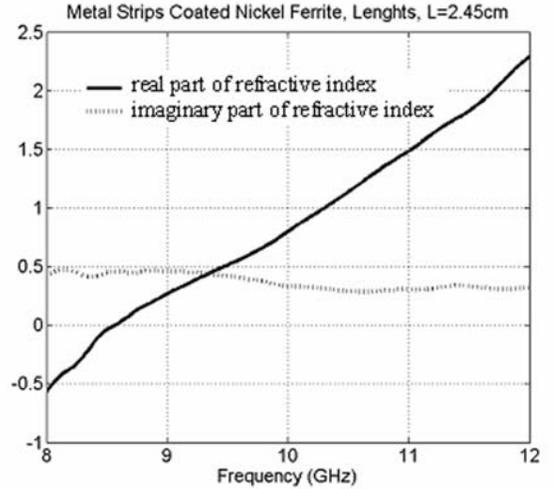

Fig.5 Measurements of complex refractive index for wire-coated nickel ferrite in waveguide.

Moreover, the positive refractive index beyond the 8.7 GHz spectra shows that the refractive index continues to increase with increases in frequency. This may be reasonable based on the theoretical calculation of the negative permeability of nickel ferrites in a 100mT field, for example. At this point, we present a brief theoretical analysis for the permeability of ferrite and the permittivity of periodic metal strips. In this waveguide measurement of this experiment, the external field $H_0$ applies along RF electric field of the $TE_{10}$ mode, that is to say the external field $H_0$ field is parallel to the wires on the ferrite disk. So, one easily obtains the effective permeability of this ferrite disk inside waveguide by follows.

$$\mu_{eff} = \frac{\mu^2 - \kappa^2}{\mu} \qquad (7)$$

where $\mu = 1 + \dfrac{f_m f_0}{f_0^2 - (f + j\Delta f_G)^2}$ , $\kappa = \dfrac{f_m (f + j\Delta f_G)}{f_0^2 - (f + j\Delta f_G)^2}$ ,

$f_0 = \gamma\sqrt{H_0(H_0 + 4\pi M_s)}$ , $f_m = \gamma 4\pi M_s$ , $\gamma = 2.8 GHz / kOe$ ,

$\Delta f_G$ is a microwave loss from the ferrite disk and it is in frequency unit which is usually described by the Gilbert damping. Perhaps it represents the intrinsic spin misalignments as well as crystal anisotropy field of ferrite disk. The $4\pi M_s$ is the saturation magnetization of the ferrite disk and $f$ is an operational frequency from the RF source, which is from the vector network analyser. A modeled permeability for a ferrites disk inside the waveguide may be seen in Fig.6.

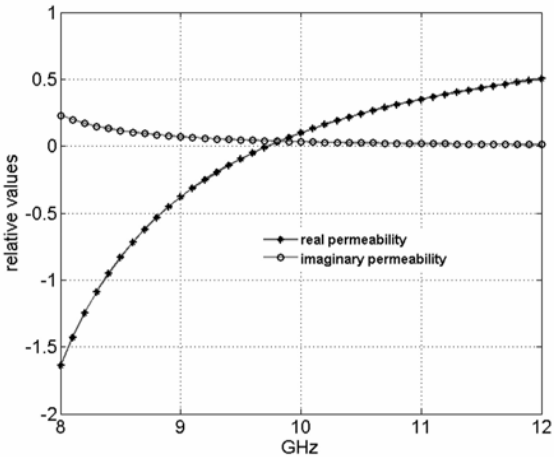

Fig.6 The theoretical calculation of complex permeability of ferrite disk inside rectangular waveguide, in which the parameters were assumed as follows: the external field, $H_0 = 100mT$ , saturation magnetization, $4\pi M_s = 1.8kOe$ , $\Delta f_G = 0.1GHz$

In order to understand the permittivity contribution from the periodic metal wires, a sample of periodic metal wires arrays were prepared on Scotch tape. The prepared sample was tested with the frequency spectrum of the X-band (see Fig.7). It is now necessary to point out that the all samples were in the same geometrical dimensions. The designed sample dimensions were the same as those of the cross section surface geometry of the rectangular waveguide. A standard rectangular waveguide of X-band was also used in this experiment. The following figure is the result of the $S_{21}$ measurement from the vector network analyser.

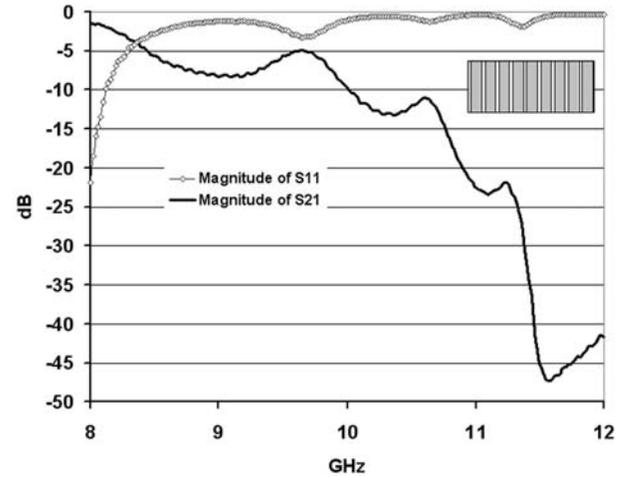

Fig. 7 Measured scattering parameters of $S_{11}$ and $S_{21}$ in dB from 8 to 12 GHz. Periodic metal wires taped on Scotch tape. Each aluminum foil tape wire width is 2 mm, the gap between the wires is 0.5 mm and the total length of the cell is 2.3 cm.

The measured $S_{11}$ and $S_{21}$ parameters (see Fig.7) for periodic metal wires applied to Scotch tape revealed that there was deeper absorption near the higher frequency end of the waveguide, while the lower frequency range allowed more transmission to pass. As such, this kind of structure of periodic metal strips provide enormously enhanced effective mass that helps to obtain the Plasmon frequency within the GHz frequency spectrum, and it appears around 12 GHz for this case (see Fig.7). It implies that the effective permittivity from periodic metal wires on Scotch tape may not be negative; but it'll be negative when they were arrayed on the ferrite disk. This part of the research may be further explained by the obtained permeability and permittivity of this wire-coated ferrite samples. Those works will be reported in the extended versions of this report.

## IV. DISCUSSION

According to the S-parameter measurements, a full band microwave isolator was achieved using ferrite samples coated with metal strips. The refractive index of the nonreciprocal circuit was derived using equation (6), which was implemented with the Cauchy-Riemann equations. Measurements of phenyloxide were taken using this novel method, and were then matched against known phenyloxide measurements. Until now, we have no problems to determine excited mediums refractive index. However, this research has been seen the positive imaginary parts of permeability and permittivity excited mediums for some time now. This kind of phenomena was also reported elsewhere, and various arguments are existed by other investigators proposals. This kind of anomaly is serious problem for classical electrodynamics analysis while others are even claiming the proof of energy conservation of electrodynamics in medium may be enough for such microwave metamaterials cases. However, it is a plan now to report this problem by deploying quantum treatment such as WKB approximation since the resonant states responsibly created the microwave metamaterials. By deploying quantum treatment, it allows the positive imaginary parts can be at large while classical dynamics does not support that kind of idea. Either case, the relevant researches will be reported in near future. Perhaps, it'll be appeared in regular transactions of Microwave Theory and Techniques then.

### ACKNOWLEDGEMENT

The research is supported by a contract from United States Army National Ground Intelligence Center.

# An Unusual Internal Anisotropy Field of Spiral Magnetized Crystal Compound of $Sr_{1.5}Ba_{0.5}Zn_2Fe_{12}O_{22}$

## Abstract

The spiral spin structured magnetic materials display important magneto-electric properties that have lead them to become known as frustrated magnets. Recently, an insulator single crystal compound of $Sr_{1.5}Ba_{0.5}Zn_2Fe_{12}O_{22}$ was also found to be a material that has magneto-electric (ME) effects. This is a rare magnetoelectric material, with its Neel temperature, at 326 K, above room temperature. As such, we measured ferromagnetic resonance (FMR) of this material at 9.5 GHz in 1.5 kOe, and the corresponding ferromagnetic resonance response showed that this compound needs - 0.5 kOe unusual internal fields to hold ferromagnetic resonance condition for this crystal compared to the ferromagnetic resonance of normal planar hexaferrites. To justify this strange internal field necessity, subsequent measurements from a SQUID (superconducting quantum interference device) magnetometer and a vibrating sample magnetometer (VSM) were also performed. Those measurements were capable of showing the saturation magnetization as $4\pi M_s = 1800$ Gauss, the internal plane anisotropy field as $H_a = 9.8$ kOe, and the line width of ferromagnetic resonance as around 50 Oe. It is interesting that this unusual internal anisotropy field helps to estimate existence of domain walls for this crystal which is similar to Bloch's wall, in which the magnetization rotates through the c-axis of this crystal; the estimated domain wall thickness is approximately 0.27 micron meter of this crystal. This kind of domain wall existence in insulator ferrite may indicate that it is a source of distinct electric polarizations, which means this crystal belongs to the magneto-electric property material.

*Index Terms*— Ferrites, ferromagnetic resonance, internal anisotropy fields, and permittivity

PACS: 75.50.Gg

## I. Introduction

The ferromagnetic crystal structure of the hexaferrite compound $Ba_{2-x}Sr_xZn_2Fe_{12}O_{22}$ is known to be helimagnetic. It belongs to a group of frustrated ferrimagnets, as named by former investigators, and recently, it has been reported to have ferroelectric properties [1, 2, and 3]. This implies that the material could potentially have a magneto-electric property that is very useful in modern technology, such as aiding in the electrical tuning for magnetic systems. Theoretical microscopic calculations show that these types of materials may have frustrated exchange interactions, and that they may induce the electric polarizations inside of crystals [3, 4, and 5]. Traditionally, this type of helical hexaferrite is also identified as a class of antiferromagnetic materials [6]. Both the past and recent reports of these materials encourage the investigation of magnetic and microwave properties through different methods. Thus, the classical measurements of FMR, VSM, and the SQUID were carried out on a single crystal hexagon disk of $Ba_{2-x}Sr_xZn_2Fe_{12}O_{22}$. In order to interpret the experimental measurements, the traditional free energy model was used to define the internal fields of this material [6, 7]. The results of this study show that an internal field of about − 0.5 kOe was needed to hold the ferromagnetic resonance condition of the normal linear microwave mode analysis. This irregularity in the behavior of the compound may be due to its unusual spin structure coming from layer interactions between the ferromagnetic and antiferromagnetic phases. According to definition, these types of helical spin systems feature exchange interactions between nearest-neighbor planes that are ferromagnetic, while those between second nearest-neighbor planes experience interactions that are antiferromagnetic [6]. Therefore, the classical microwave mode analysis, or Kittel mode, of this system calls for a previously undiscovered internal field to explain experimental measurements. As such, an anisotropy field, namely an internal anisotropic field or demagnetizing damping field, is considered to fill this role.

## II. Ferromagnetic resonance condition for the crystal composition of spin spirality in easy plane magnetization ⊥ to the c-axis

The composite of $(Sr,Ba)Zn_2Fe_{12}O_{22}$ was studied in the early 1960's for its helical spin configuration, and it was because of the identified internal anisotropy fields in this composite that further analysis was also carried out to understand the internal anisotropy fields for other such composites. Until now, there has been no convincing argument that explains the force that induces the helical spin alignments that disturb planar magnetizations in an anisotropy field with high magnitude (on the order of one Tesla). Although there were already several practical arguments in existence for the reason behind the helimagnetic status of this composition, here the free energy terms are rewritten and presented for the helical hexaferrite crystal disk in this paper. Supposing one can follow the classical approach for the planar helical hexagonal crystal disk's free energy [7], it is expanded using a power series of cosine as follows:

$$F_h = -M_s H \sin \vartheta \cos \varphi$$
$$+ \frac{1}{2}(N_x M_s^2 \sin^2 \vartheta \cos^2 \varphi + N_y M_s^2 \sin^2 \vartheta \sin^2 \varphi \qquad (1)$$
$$+ N_z M_s \cos^2 \vartheta) - K_g \cos^2 \vartheta - K_a \cos \varphi - K_\varphi \cos 6\varphi$$

In equation (1), $\vartheta$ [7] polar angle, $\varphi$ [7] azimuth angle, and $N_x$, $N_y$, and $N_z$ are the demagnetizing factors, $K_a$ is introduced to this crystal in this paper which represents a force that causes the spiral magnetization in this system. It may be understood that it is an exchange interaction between ferromagnetic and anitferromagnetic layers in this system. One proposes that the hexagonal crystal disk of this composite is found to have the following layer sequence: (AFM) (FM) (FM) (AFM) (FM) (FM) (AFM) along the c-axis. If so the potential domain walls configuration inside crystal behave like Bloch walls. It implies that the magnetization rotates through the c-axis of this compound. This kind of macroscopic ordering generates a fairly strong repulsion force between the ferromagnetic layers, a force that is strong enough to cause the level of spiral magnetization that is seen in the composite. In such a scenario, the internal anisotropy filed may be estimated [8]. The effective anisotropy fields may be defined as follows [9]:

$$\frac{\partial F_h}{\partial \vartheta} = 2K_g \cos \vartheta \sin \vartheta = M_s H_g \sin \vartheta$$
$$\vartheta \approx 0 \qquad (2)$$
$$H_g = \frac{2K_g}{M_s}$$

$$\frac{\partial F_h}{\partial \varphi} = K_a \sin \varphi + 6K_\varphi \sin 6\varphi$$
$$= M_s H^{eff} \sin \varphi = M_s (H_\varphi + H_a) \sin \varphi$$
$$\varphi \approx 0 \qquad (3)$$
$$H_\varphi \approx \frac{6K_\varphi}{M_s} \frac{\sin 6\varphi}{\sin \varphi} = \frac{36K_\varphi}{M_s}$$
$$H_a \approx \frac{K_a}{M_s}$$

Now the ferromagnetic resonance condition is presented. For this special case, we assume the c-axis of the crystal is positioned along the z-axis, the DC magnetic field strength, H is along the y-axis, and the RF magnetic field (TEM) propagates along the y-axis. The effective static and microwave fields inside crystal disk may be represented as follows:

$$H_i = H$$
$$h_i^x = h_x - \frac{H^{eff}}{M_s} m_x \qquad (4)$$
$$h_i^z = h_z - N_z m_z - \frac{H_g}{M_s} m_z - \frac{H^{eff}}{M_s} m_z$$

Using the above equations (2), (3), and (4) with the motion equation for magnetization [6, 9], one obtains the ferromagnetic resonance condition:

$$\frac{\omega_h}{\gamma} \approx \sqrt{(H + H_\varphi + H_a)(H + H_\varphi + H_a + H_g + 2\pi M_s)} \quad (5)$$

In the derivation process for equation (5), approximations for the special hexagon disk were used. That is, the aspect ratios that are responsible for the demagnetization of this single crystalline hexaferrite disk were measured to be the following:

$$\left| N_y - N_x \right| \approx 4\pi \left| \frac{h}{a} - \frac{h}{b} \right| \approx 0 \text{ and } N_z - N_x \approx 2\pi .$$

Clearly, if the demagnetization of magnetic materials induced spiral structure, then all magnetic materials would be capable of forming such a spiral configuration. Demagnetization could therefore not be the means by which the spiral structure is formed.

### III. The measurements of A crystal of planar helical hexaferrite

The Bell Laboratories, Kimura provided a platelet crystal of $Sr_{1.5}Ba_{0.5}Zn_2Fe_{12}O_{22}$ [1]. The properties of the crystal were mass m = 4.9mg and geometrical dimensions $a \approx 1.5$mm, $b \approx 1.3$mm and h = 0.73mm. The Neel temperature for this composition was reported as $T_N = 326$K and magnetically induced ferroelectricity was up to 130 K, an observable phenomenon perpendicular to the c-axis of the crystal [1]. In the study, the easy magnetization plane (external field perpendicular to the c-axis) and hard magnetization plane (external field parallel to the c-axis) temperature-dependent magnetization phase diagrams were measured by SQUID at various temperatures, ranging from 5 Kelvin to 300 Kelvin (see Fig.1 and Fig.2).

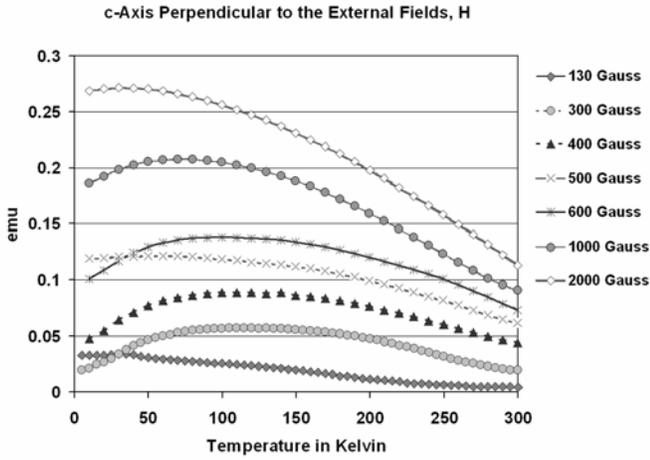

Fig.8 Temperature dependence of saturation magnetization in the plane by SQUID measurements at various temperatures, where the external fields were applied perpendicular to the c-axis of the crystal disk.

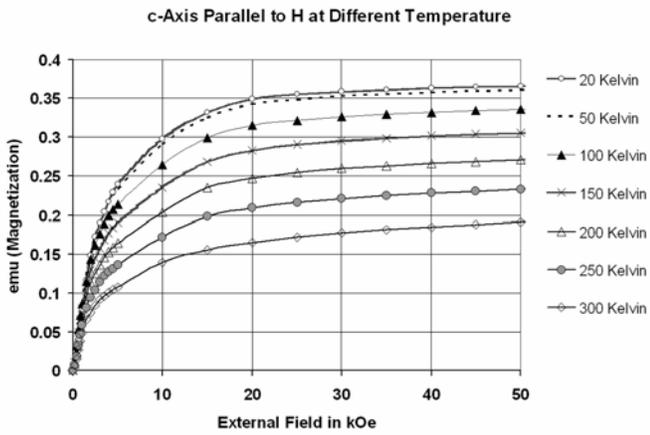

Fig. 9 Temperature dependence of saturation magnetization in the plane by SQUID measurements at various temperatures, where the external fields were applied parallel to the c-axis of the crystal disk

In Fig. 1, the basal plane SQUID measurement shows that the temperature-dependent magnetizations were quite non-linear in behavior around the temperature of 100 Kelvin. However, in Fig. 2 there are no specific differences in the linearity of magnetizations at different temperatures. According to Figures 1 and 2, it appears that the force causing spiral magnetization may be laid in the plane of magnetization, which may be perpendicular to the c-axis of the hexaferrite disk. If so, it further supports the existence of the unusual internal anisotropy, field $H_a$ .In equation (1) and (3) it was written in terms of azimuth angle, and $\varphi$ was expressed in the physical sense. In addition, the SQUID measurement helps determine the saturation magnetization in the plane of this crystalline disk at room temperature. It was measured to be as $4\pi M_s = 1800$ Gauss (see Fig.1). In order to further understand internal anisotropy fields and saturation magnetizations at room temperature, we have also made the measurements of VSM and FMR for this crystalline disk. The VSM measurement confirms the SQUID measurement in terms of saturation magnetization, and it also shows that the in-plane anisotropy field was around 9.8kOe (see Fig.3). The FMR measurement further confirms the in-plane internal anisotropy field measurement by VSM and it helps to determine the internal anisotropy field of this material as well (see Fig.4).

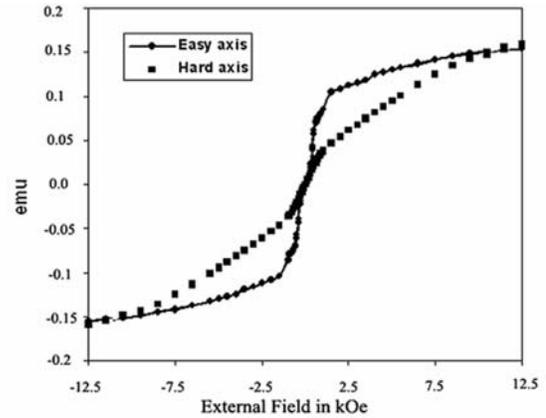

Fig. 3 VSM measurements of this single crystalline for the hard and easy axes of the magnetizations at room temperature

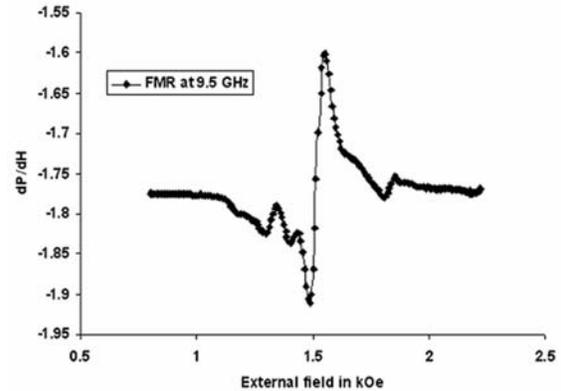

Fig. 4 FMR measurements at 9.5GHz, A -0.5 kOe anisotropy field was needed to hold the normal ferromagnetic resonance condition point, which occurred at an external field of 1.5 kOe for the crystalline disk

In general cases, the basal plane anisotropy field of planar hexaferrites, $H_\varphi$, are very small and within the order of Oe. This allows the ferromagnetic resonance condition of this crystalline sample may to be simplified to the following:

$$\frac{\omega_h}{\gamma} \approx \sqrt{(H+H_a)(H+H_a+H_g+2\pi M_s)} \qquad (6)$$

In order to interpret the above ferromagnetic resonance of this sample at 9.5 GHz in an external field of H = 1.5 kOe, the estimated unusual anisotropic field magnitude, $H_a$, may be found as follows, $H_a = K_a / M_s \approx -0.5 kOe$. It is our understanding that this force may be understood by Malozemoff findings [8]. As such this study finds it as follows, $K_a = \frac{2\sqrt{A_{ex}K_g}}{t_h}$. Usually, the exchange stiffness constant by follows, $A_{ex} \approx 1.5 \times 10^{-6} \, erg/cm$ and this study finds the crystallography anisotropy energy toward c-axis by follows, $K_g \approx 0.75 \times 10^6 \, erg/cm^3$. In order to generate the unusual 500 Oe anisotropy field, the estimated domain wall thickness along the c- axis has to be $t_h = 0.27 \mu m$ if Malozemoff finding is acceptable to readers. This is an amazingly good data to be accepted for this crystal and this study suggests taking credit from the research [8]. According to historical investigations [10, 11, 12], one could assume one of several possibilities for the force that causes the spiral structure of the magnetization of this crystal. The first is the possibility of oscillating states from lattice parameters due to the discrete magnetizations seen in the easy magnetizations plane [10]. The next suggests that there may be magnetostriction along the c-axis due to the easy magnetization hardness that was observed along the c-axis [11]. The last possibility is that the larger negativity of this field could be the frustration factor source from the exchange interactions between ferromagnetic and antiferromagnetic layers. Considering macroscopic reasons, the force may also be the result of interaction between moments in different micromagnetic domains due to the quasi-continuous spiral magnetization structure of the crystalline disk. The negativity of this field may also be due to the fact that the interaction between micromagnetic domains (postulated multi-domains for negative anisotropy) may be such that they align themselves into internal energy minimizing directions, like a demagnetizer, along the c-axis [13]. According to the findings of this research and the model of the free energy for this crystalline disk from this paper, this strange internal field may be due to domain walls between antiferromagetic and ferromagnetic layer sequences was imagined for this disk, that is: (AFM) (FM) (FM) (AFM) (FM) (FM) (AFM) is the earlier stated sequence of the disk. Most likely the domain walls formation similar Bloch's wall along c-axis and the magnetization rotate through the c-axis of this crystal. After all, this extra field is needed to hold this crystalline material's FMR measurement at 9.5GHz for the linear microwave mode (Kittel's mode) analysis of this crystalline sample. Based on those suggestions and free energy terms for the crystal from this paper, as well as the level of the strange internal field that was observed, the ferromagnetic resonance in Fig.4 shows that the repelling force between the ferromagnetic layers of the crystal would be able to cause such spiral magnetization in the easy magnetized plane. In contrast to that, the other possibilities of causing such a result are ineffectual, based on their magnitudes, in causing such spiral magnetization for this hexaferrite disk.

## IV. Domain wall's application of this crystal

If this crystal could be induced to the electric dipoles when applying the magnetic field perpendicular to the c-axis [1], it implies that the crystal domain walls lose its stable state, which is perpendicular to the external magnetic field. In such circumstances the oscillation forces are as follows:

$$F = -K_m \cdot x \qquad (7)$$

The $K_m$ is the restoring force of displacement of dipoles from domain wall oscillation (it may be the force which was seen in the ferromagnetic resonance in Fig. 4) and x is the displacement vector. The field's negativity is then inducing demagnetizing behavior from domain walls against external magnetic field exertion. An applied appropriate periodic external pulse field, $H_d$, may maintain the oscillatory states in the crystal. If so one obtains the following:

$$m\dot{v} + K_m x = -evH_d$$

$$j\omega v + \frac{K_m}{m} x = -\frac{e}{m} vH_d \qquad (8)$$

$$v = \frac{-\frac{K_m}{m} x}{j\omega + \gamma H_d} = j\frac{\omega_{ex}^2 x}{\omega - j\omega_d}$$

Where $v$ is the velocity of oscillatory waves caused by disturbing external fields toward domain wall, $\gamma$

is the gryromagnetic ratio, $\omega_d = \gamma H_d$ is resonator frequency from the pulse field, m is the effective mass of vibrating domain walls of the crystal and $\omega_{ex} = \sqrt{\frac{K_m}{m}}$ is intrinsic frequency. It is obvious now that there may be an effective conductive current through the domain walls along the c-axis, which is related to the term of these vibrating states. This may be written as follows:

$$J_v = -Nev = -j\frac{\omega_{ex}^2 Nex}{\omega - j\omega_d} = -j\frac{\omega_{ex}^2 P}{\omega - j\omega_d} \qquad (9)$$

For checking to see this crystal system's effective permittivity, one may rewrite the Maxwell equation, which may be as follows:

$$\nabla \times \vec{H} = \frac{\partial \vec{D}}{\partial t} + \vec{J}_v$$

$$= j\omega\varepsilon\vec{E} - j\frac{\omega_{ex}^2 \vec{P}}{\omega - j\omega_d} \qquad (10)$$

$$= j\omega\varepsilon\vec{E}(1 - \frac{\omega_{ex}^2}{\omega(\omega - j\omega_d)}\frac{\vec{P}}{\varepsilon\vec{E}})$$

From the above formulas one can conclude that the effective permittivity of this crystal system is as follows.

$$\varepsilon_{eff} = \varepsilon(1 - \frac{\omega_{ex}^2}{\omega(\omega - j\omega_d)}\frac{P}{\varepsilon E}) \qquad (11)$$

Equation (11) reveals an important potential application of the effective permittivity in future technology, that being that the permittivity of this crystal system can be tuned by external electric as well as magnetic fields if the crystal has the ferroelectric property P [1]. It urges one to believe that a possible negative refractive index may also be possible in insulator ferrites such as $Sr_{1.5}Ba_{0.5}Zn_2Fe_{12}O_{22}$.

## V. Conclusion

In this article, an unusual internal anisotropic field, $H_a$, was presented, which has a negative value of -0.5 kOe. The basal plane anisotropy field ($H_\varphi$) was regarded as negligible in this study due to the presence of large anisotropy fields, for example, $H_a \gg H_\varphi$ and $H_e \gg H_\varphi$, so it would not be used in fulfilling the ferromagnetic resonance condition for the planar hexaferrite compound of this single crystalline disk. Beyond that, a reiterated methodology is for deriving the ferromagnetic resonance condition for hexaferrites using Kittel's torque technique is presented. This methodology easily applies to the case of single helical crystalline hexaferrite disk as well. Both static and dynamic considerations on the system are accounted for, making it more useful than Smit and Beljers' ferromagnetic resonance condition derivation, as they defined the resonance condition only on pure static magnetization systems. The origin of the negative anisotropic field was briefly discussed in this paper for further study purposes. Moreover, an important possible application for this material was proposed through reasonable formulas.

### ACKNOWLEDGEMENT

This study is partially supported by a contract from the US Army National Ground Intelligence Center.

# Mahmut Obol, Ph.D.

EMAIL: mahmut.obol@gmail.com

## CAREER GOALS

To work in ceramics and biophysics R&D sector developing magnetic functional materials for next generation ceramics technologies as well as their applications into biophysics technologies.

## EDUCATION

**Ph.D.**  Electrical and Computer Engineering, Northeastern University (Boston, MA)    2004
**M.S.**   Physics, Northeastern University (Boston, MA)    2000
**M.S.**   Physics, Institute of Atomic Energy of China (Beijing, China)    1991
**B.S.**   Physics, Xin Jiang University (Urumqi, China)    1988

## WORK EXPERIENCE AND HANDS ON TECNIQUES

1) Magnetic and dielectric materials microwave characterizations in waveguide. This technique is in state of determine materials refractive index, impedance, permeability and permittivity etc without using any guess parameters and this technique is free any divergence from the waveguides at all. This technique has also several professional tactics to determine normal material, resonant material, and metamaterials microwave characterizations in terms of refractive index, impedance, permeability and permittivity.

2) Single and poly crystalline ferrites growth by flux melt and sintering methods and their crystal axis orientation technique. This technique is in state of grow and produce numerous single and poly hexaferrites which includes magneto-electric strontium rich planar hexaferrite. The strontium rich planar hexaferrite is going to have potential microwave applications such as sensors.

3) Coaxial probe technique for sensing complex permittivity and conductivity of liquid and semi liquid materials such as biological tissues. This is fast detecting technique for non magnetic materials microwave properties.

4) Microstripline measurement technique in terms of detecting permeability and permittivity of oxides. This technique is in state to detect materials microwave properties from 1 GHz to 4GHz spectra where the waveguide technique is challenged by its physical dimension. This technique's capability is great to compensate covering such frequencies spectra.

## INSTRUMENTATION AND COMPUTER SKILLS

- Deploying microwave measurements on waveguide, coaxial probe, cavity and microstripline with Agilent's Vector Network Analyzer at broad GHz frequencies spectra.
- Characterized magnetic and structural properties of materials using XRD, SQUID, SEM, VSM and FMR.
- Professional computational tool is for me MATLAB.
- Designed passive devices using electromagnetic simulation tools including Sonnet[TM].
- Familiar publishing tools are Microsoft Office and Photoshop etc.

## PERSONAL AND SPIRIT

To devote in fast faced work spirited environments, admire people but can't stand talks too much, and practicing positive progressive attitude in research and development. Respect individual and team efforts.

## LANGUAGES

English, Chinese, and Uyghur (native)

## PROOFS OF HANDS ON TECHNIQUES

The hands on techniques will be presented for the interested research sectors and companies by practice.

## EMPLOYMENT

| | |
|---|---|
| 2006-2010 | Research Staff, Tufts University (Medford, MA) |
| 2005-2006 | Associate Professor, Xinjiang Normal University (Urumqi, China) |
| 2000-2004 | Research Assistant, Northeastern University (Boston, MA) |
| 1998- 2000 | Teaching Assistant, Physics Department, Northeastern University (Boston, MA) |
| 1991-1998 | Lecturer, Physics Department, Xin Jiang University (Urumqi, China) |

## TEACHING CAPABILITY

Quantum mechanics, Electrodynamics, Classical Mechanics, and Circuits for Undergrade and Graduate Levels